\documentclass[onecolumn,amsmath,amssymb,aps,showpacs,superscriptaddress]{revtex4}

\usepackage[dvips]{graphicx}
\usepackage[dvips]{color}
\usepackage{hyperref,breakurl,amsmath,amssymb,pst-node,eepic}
\usepackage{graphicx}
\usepackage{dcolumn}
\usepackage{bm}
\usepackage{hyperref}

\begin{document}

\title{The fine structure of spectral properties for random correlation matrices: an application to financial markets}

\author{Giacomo Livan}
\email{giacomo.livan@pv.infn.it}
\affiliation{Dipartimento di Fisica Nucleare e Teorica, Universit\`a degli Studi di Pavia, Via Bassi 6, 27100 Pavia, Italy}
\affiliation{Istituto Nazionale di Fisica Nucleare, Sezione di Pavia, Via Bassi 6, 27100 Pavia, Italy}
\author{Simone Alfarano}
\email{alfarano@eco.uji.es}
\affiliation{Departament d'Economia, Universitat Jaume I, Campus del Riu Sec, 12071 Castell\'on, Spain}
\author{Enrico Scalas}
\email{enrico.scalas@mfn.unipmn.it}
\homepage{www.mfn.unipmn.it/~scalas}
\affiliation{Dipartimento di Scienze e Tecnologie Avanzate,
Laboratorio sui Sistemi Complessi, Universit\`a del Piemonte Orientale ``Amedeo Avogadro'',
Viale T. Michel 11, 15121 Alessandria, Italy}

\date{\today}

\begin{abstract}
We study some properties of eigenvalue spectra of financial correlation matrices. In particular, we investigate the nature of the large eigenvalue bulks which are observed empirically, and which have often been regarded as a consequence of the supposedly large amount of noise contained in financial data. We challenge this common knowledge by acting on the empirical correlation matrices of two data sets with a filtering procedure which highlights some of the cluster structure they contain, and we analyze the consequences of such filtering on eigenvalue spectra. We show that empirically observed eigenvalue bulks emerge as superpositions of smaller structures, which in turn emerge as a consequence of cross-correlations between stocks. We interpret and corroborate these findings in terms of factor models, and and we compare empirical spectra to those predicted by Random Matrix Theory for such models.
\end{abstract}


\maketitle

\section{Introduction}

In Physics, Random Matrix Theory (RMT) is mainly used to model systems of particles interacting according to unknown laws. This is particularly handy for studying energy levels of complex systems such as heavy nuclei and mesoscopic systems. In such cases, the Hamiltonian operator can be conveniently described by a random matrix featuring some suitable symmetry properties. In particular, two matrix ensembles have been commonly used: the Gaussian Orthogonal Ensemble of real symmetric random matrices, and the Gaussian Unitary Ensemble of Hermitian random matrices \cite{Wigner, Mehta}. In both these cases, for proper normalization of matrix elements, the asymptotic statistical properties of the eigenvalues follow the so-called semicircle law:

\begin{equation}
\rho(\lambda) = \frac{1}{2 \pi} \sqrt{4 - \lambda^2},
\end{equation}
where $\rho(\lambda)$ is the marginal probability density function of the eigenvalues. Until recent years, physicists often neglected the study of random correlation matrices, even though they find applications in very diverse fields ranging from biology to econometrics. For this reason, applied mathematicians have studied such objects since the 1920s \cite{Wishart}. The asymptotic eigenvalue statistics in this case is given by the Mar\v cenko-Pastur distribution \cite{Marcenko}, which will be extensively discussed in the following sections. Since the late 1990s, thanks to the growing interest in financial markets as prototypes of complex systems, physicists started working on random correlation matrices \cite{Laloux, Plerou}, and this will be the subject of this paper as well. \\
We consider a set of $N$ stocks whose spot price at time $t$ we denote as $S_i(t)$, $i = 1, \ldots, N$. Let $t_1, \ldots, t_{T+1}$ be $T+1$ equally spaced time instants, then we introduce the corresponding log-returns

\begin{equation} \label{returns}
r_{i,j+1} \doteq  \log \frac{S_i(t_{j+1})}{S_i(t_j)};
\end{equation}
typically, one can think of the $t_i$ as days. This notation is a little redundant, and we can simply denote time steps as $j = 1, \ldots, T+1$. Now, we can assume that the $T$ recorded log-return values are  realizations of $N \times T$ random variables $R_j^{(i)}$, so that we globally end up with $NT$ observations $r_{ij}$, $i = 1, \ldots, N$, $j = 1, \ldots, T$. Equivalently, the vector

\begin{equation}
\mathbf{r}_i \doteq ( r_{i,1} , \ldots, r_{i,T} )
\end{equation}
containing all the observations of the $i$th asset returns, can be seen as a realization of a vector random variable $\mathbf{R}^{(i)}$. Such a framework is fully characterized by finite probability distributions \cite{Kolmogorov, Billingsley}:

\begin{equation} \label{finiteprob}
\mathbb{P} \left (R_1^{(1)} \in A_1^{(1)}, \ldots, R_T^{(1)} \in A_T^{(1)}; \ldots ; R_1^{(N)} \in A_1^{(N)}, \ldots, R_T^{(N)} \in A_T^{(N)} \right ) =
\mathbb{P} \left ( \mathbf{R}^{(1)} \in B^{(1)}; \ldots; \mathbf{R}^{(N)} \in B^{(N)} \right )
\end{equation}
where $A_j^{(i)} \in \mathbb{R} \ \forall i,j$ and $B^{(i)} \in \mathbb{R}^T \ \forall i$. Depending on the choice of the random variables $\mathbf{R}^{(i)}$s, such a picture allows for a huge variety of possible descriptions of the stochastic dynamics of financial data. Most simply, a standard assumption, according to which the log-returns are described by uncorrelated Gaussian processes ($(r_{1,i}, \ldots, r_{N,i}) \sim \mathcal{N}(0,\mathbf{1}_N)$, where $\mathbf{1}_N$ represents the $N \times N$ identity matrix), could be adopted. However, as is well known \cite{Campbell}, correlations often play a major role, and a realistic description of financial markets should by no means neglect them. Still, a Gaussian framework can be retained by observing that a set of zero-mean correlated Gaussian numbers generated by a stationary stochastic process is completely characterized by its expectation value vector $\boldsymbol{\mu}$ and covariance matrix $\boldsymbol{\mathcal{E}}$:

\begin{equation} \label{cov}
\mathcal{E}_{ij,kl} = \mathbb{E} \left [ r_{ij} r_{kl} \right ].
\end{equation}
Following \cite{Burda2010}, in this paper we shall simplify this structure to the assumption that cross-correlations between assets and auto-correlations in time factorize:

\begin{equation} \label{factorcov}
\mathcal{E}_{ij,kl} = C_{ik} A_{jl}.
\end{equation}
In the above equation $C_{ik}$ ($A_{jl}$) represents an element of a $N \times N$ ($T \times T$) positive-definite symmetric matrix $\mathbf{C}$ ($\mathbf{A}$). We shall keep this same kind of notation, \emph{i.e.} denoting matrices by bold letters and the corresponding matrix elements by the same non-bold letters, throughout the rest of the paper. We shall assume the $\mathbf{C}$ matrix of cross-correlations to be constant over time. Also, most importantly, we shall neglect all possible correlations in time by assuming $\mathbf{A} = \mathbf{1}_T$:

\begin{equation} \label{notimecov}
\mathcal{E}_{ij,kl} = C_{ik} \delta_{jl}.
\end{equation}
This last assumption is well motivated both from an empirical viewpoint \cite{Campbell} and a theoretical one, since asset returns can be shown not to display auto-correlations whenever assets are assumed to be described by a sub-martingale. As a matter of fact, from the sub-martingale property one can show that

\begin{equation}
\mathbb{E}[r_{i,j} r_{i,k}] \sim \mathbb{E}[\widetilde{r}_{i,j} \widetilde{r}_{i,k}] = 0
\end{equation}
where
\begin{equation} \label{subm3}
\widetilde{r}_{i,j} = \frac{S_i(t_{j+1}) - S_i(t_j)}{S_i(t_j)} \sim r_{i,j}
\end{equation}
assuming no dividend payment in the period and $S_i(t_{j+1}) - S_i(t_j) \ll S_i(t_j)$. This relation means that returns are uncorrelated (not necessarily independent) random variables, as can be empirically verified \cite{Campbell}. In the following, we shall always assume the previously mentioned condition ($S_i(t_{j+1}) - S_i(t_j) \ll S_i(t_j)$) to be fulfilled, thus allowing to identify log-returns and returns (as in equation 
\eqref{subm3}). \\
The Gaussian probability measure leading to the correlation structure \eqref{notimecov} can be shown to be

\begin{equation} \label{probmeas}
P(\mathbf{R}) \mathrm{D}\mathbf{R} = \frac{1}{(2\pi)^{NT/2} (\mathrm{det} \mathbf{C})^{T/2}}
\exp \left ( - \frac{1}{2} \mathrm{Tr} \mathbf{R}^{\mathrm{T}} \mathbf{C}^{-1} \mathbf{R} \right ) \mathrm{D}\mathbf{R}
\end{equation}
where $\mathbf{R}$ is a rectangular $N \times T$ matrix containing all of the returns observations ($R_{ij} = r_{ij}$), while $\mathrm{D} \mathbf{R} \doteq \prod_{i=1}^N \prod_{j=1}^T \mathrm{d} R_{ij}$ is the flat integration measure over matrix elements. \\
Being symmetric, the $\mathbf{C}$ matrix in \eqref{probmeas} is made of $N(N+1)/2$ independent entries. Now, the typical challenge to be faced in many multivariate analysis problems is to estimate these numbers from $N$ time series of $T$ observations, \emph{i.e.} $NT$ data points. When such data are collected in a $N \times T$ matrix $\mathbf{R}$ as in \eqref{probmeas}, then a standard estimator for $\mathbf{C}$ is given by the matrix $\mathbf{c} \doteq \mathbf{R} \mathbf{R}^{\mathrm{T}} / T$. In other words, an estimator for the matrix element $C_{ij}$ in $\mathbf{C}$ is given by

\begin{equation} \label{Pearson}
c_{ij} = \frac{1}{T} \sum_{t=1}^T R_{it} R_{jt},
\end{equation}
which is the well-known Pearson estimator for large $T$ (for small values of $T$ the $1/T$ factor would need to be replaced by $1/(T-1)$). Of course, the $c_{ij}$s are a noise-dressed representation of the $C_{ij}$s. As a matter of fact, even though the random variables in $\mathbf{R}$ were exactly described by the probability distribution in \eqref{probmeas} (\emph{i.e.} by the correlation matrix $\mathbf{C}$), the finiteness of the data sample under study would anyway cause the $c_{ij}s$ to deviate, on average, from their ``true'' counterparts $C_{ij}s$. As it is intuitively clear, the two will become closer as more observations are collected, \emph{i.e.} as $T \rightarrow \infty$, or equivalently as $q \rightarrow 0$, where $q$ is the so called ``rectangularity ratio'':

\begin{equation} \label{rectratio}
q \doteq \frac{N}{T}.
\end{equation}
However, realistic situations in financial practice typically involve large numbers of variables and similarly large numbers of observations. Ideally, this is not far from a ``thermodynamic limit'' situation in which

\begin{equation} \label{thermlim}
N,T \rightarrow \infty \ , \ \ \ \mathrm{with} \ \ \ \frac{N}{T} = q = \mathrm{constant}.
\end{equation}
Remarkably, this is precisely the regime under which some powerful RMT results are valid \cite{Marcenko}. In particular, this is the limit under which it is possible to make analytical statements about the relation between the eigenvalue spectra of the theoretical covariance matrix $\mathbf{C}$ and its estimator \eqref{Pearson} \cite{Silverstein, Burda2004_1, Burda2004_2}. We shall exploit those results in the following sections. \\
A very general class of models fulfilling \eqref{notimecov} is the one of the so called factor models \cite{Markowitz, Sharpe, Lillo, Marsili1, Marsili2}. Such models aim at describing the time evolution of each asset in terms of a few ``driving forces'', or factors, which typically describe the impact that a market sector or the whole market itself have on a given asset. In a $K$ factors model, the time evolution of asset returns is given by

\begin{equation} \label{factormodel}
r_{it} = r_i(t) = \sum_{j=1}^K g_i^{(j)} m_j(t) + g_i^{(0)} \epsilon_i(t),
\end{equation}
where $g_i^{(j)}$ and $g_i^{(0)}$ are constant parameters, whereas $m_j$ and $\epsilon_i$ are independent and identically distributed normal random variables. We shall assume these latter to be normalized as follows:

\begin{figure}
\begin{center}
\includegraphics[width=2.9 in, height=2.0 in]{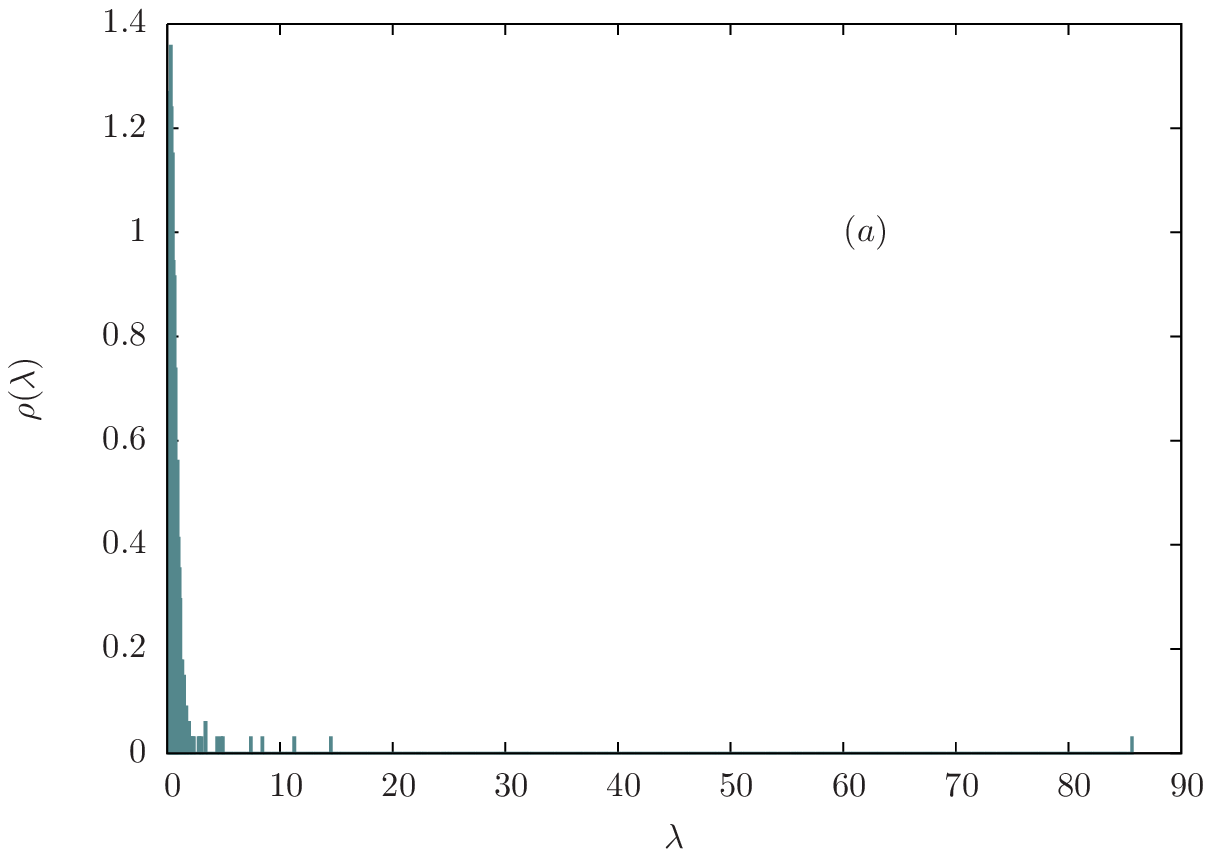} 
\hspace{0.1 in} 
\includegraphics[width=2.9 in, height= 2.0 in]{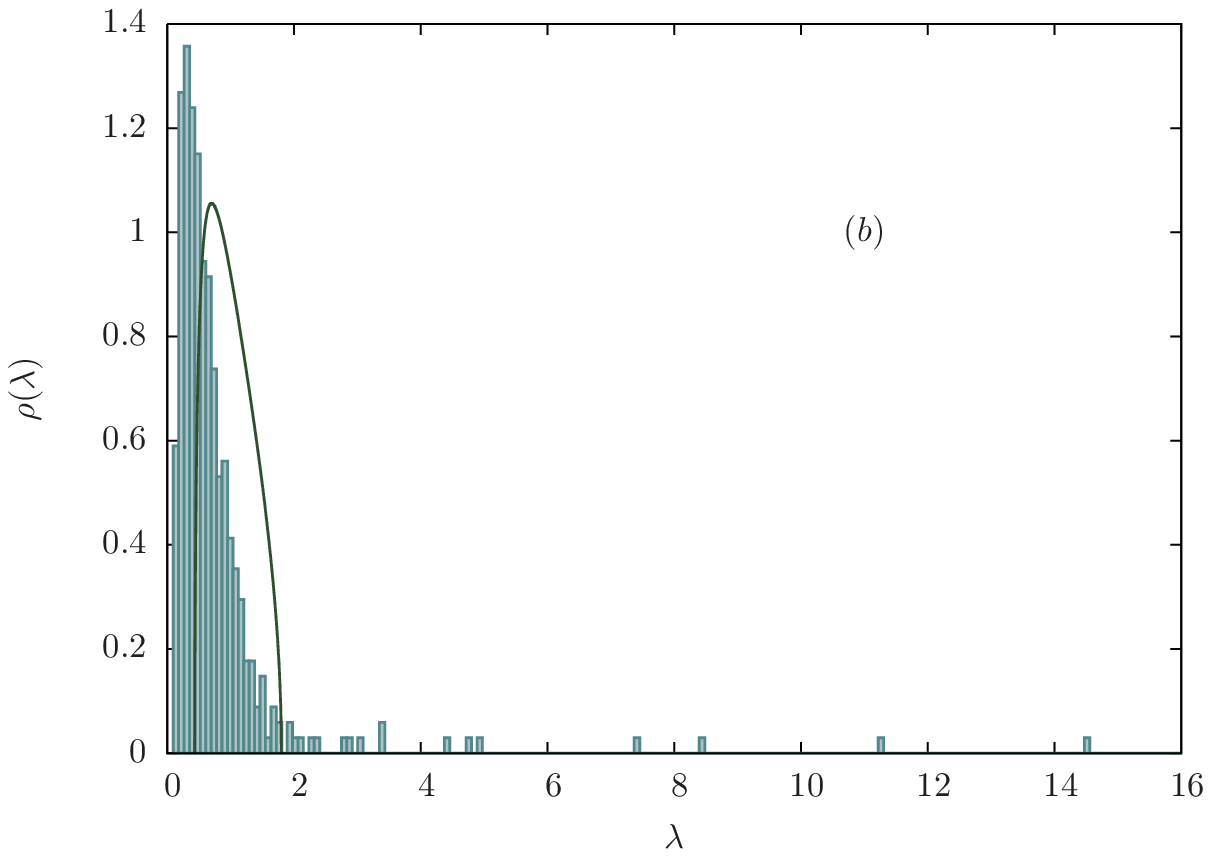} 
\vspace{0.1 in}
\includegraphics[width=2.9 in, height= 2.0 in]{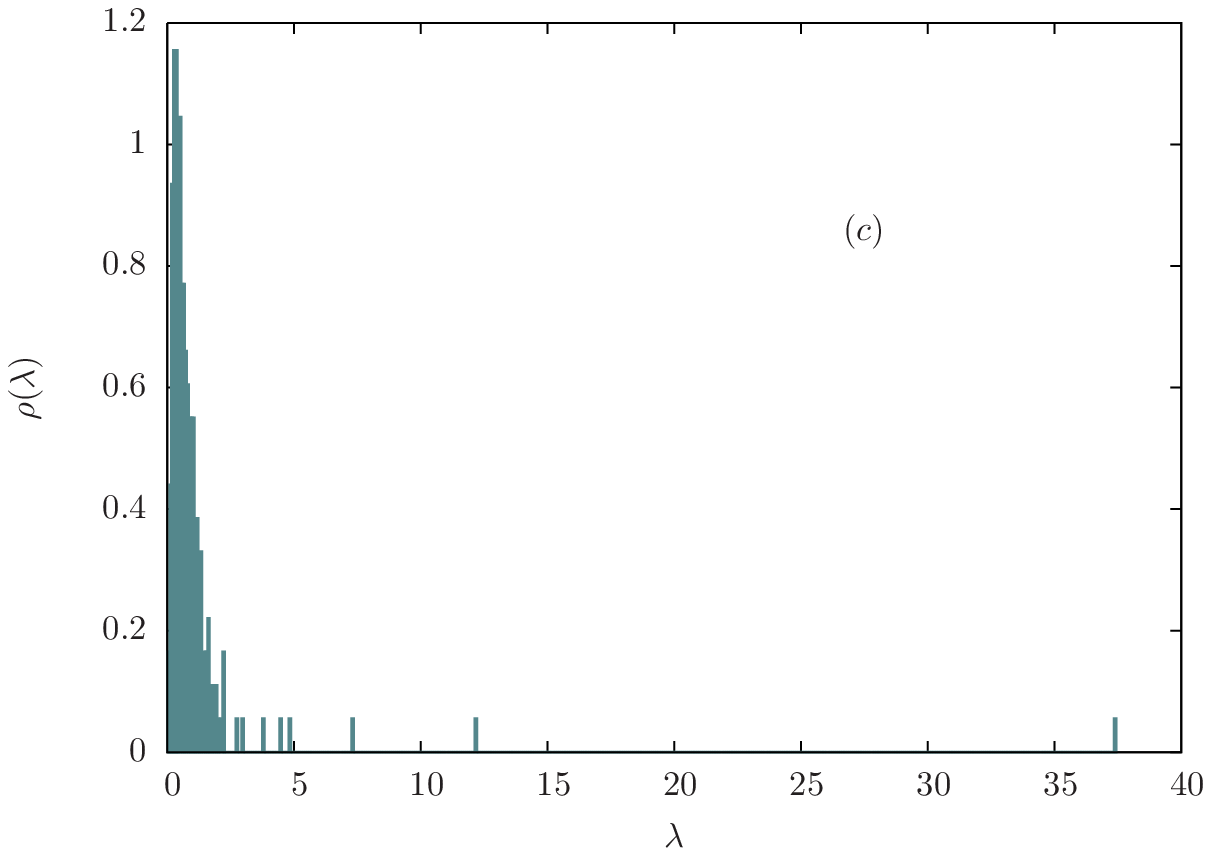}
\hspace{0.1 in} 
\includegraphics[width=2.9 in, height= 2.0 in]{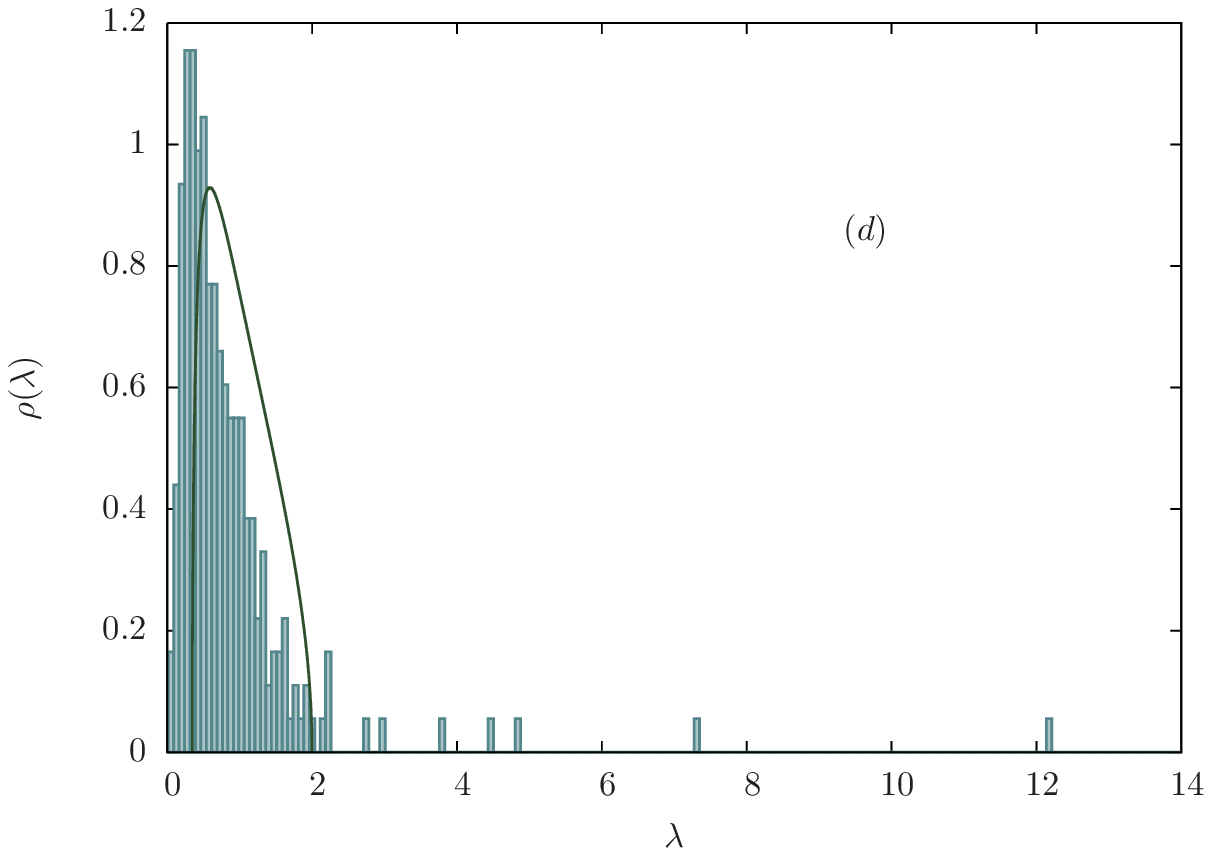} 
\caption{(a) Empirical eigenvalue density of the covariance matrix for $T = 3400$ daily returns of $N = 396$ assets belonging the S\&P500 Index over the years $1996-2009$. (b) The same density as in (a) without the largest eigenvalue. (c) Eigenvalue density for $T = 1423$ daily returns of $N = 243$ assets belonging to the FTSE$350$ Index over the years $2005-2010$. (d) The same density as in (c) without the largest eigenvalue. All figures were produced with standardized data. In (b) and (d) the Mar\v cenko-Pastur distributions for the corresponding values of $q = N/T$ are also plotted.} \label{SP_FTSE_spectra}
\end{center}
\end{figure}
\begin{eqnarray} \label{normalizations}
\mathbb{E} \left [ m_i (t) \right ] &=& \mathbb{E} \left [ \epsilon_i (t) \right ] = 0 \\ \nonumber
\mathbb{E} \left [ m_i (t) m_j (t^{\prime}) \right ] &=& \mathbb{E} \left [ \epsilon_i (t) \epsilon_j (t^{\prime}) \right ] = \delta_{ij} \delta_{t t^{\prime}} \\ \nonumber
\mathbb{E} \left [ m_i(t) \epsilon_j(t^{\prime}) \right ] &=& 0.
\end{eqnarray}
In the next section we shall specialize the model in \eqref{factormodel} to a particular case. However, in a very general fashion, factor models have proven to be able to reproduce, at least qualitatively, some relevant features of empirical covariance matrix eigenvalue spectra. \\
The general appearance of the return covariance matrix eigenvalue spectrum of a given number of assets (for zero mean and unit standard deviation data) is the one depicted in Figure \ref{SP_FTSE_spectra} for the log-returns of the daily prices for the assets composing the S$\&$P500 and FTSE350 Indices. Three main features are clearly visible: a large bulk close to zero, a number of larger eigenvalues ``leaking out'' of such bulk, and a much larger and isolated eigenvalue. Since the pioneering works \cite{Laloux, Plerou}, RMT has become a standard tool to analyze these macroscopic features. More specifically, the aforementioned eigenvalue bulk has mostly been identified with the Mar\v cenko-Pastur distribution \cite{Marcenko}, \emph{i.e.} the limiting eigenvalue marginal probability density for the (already introduced) matrix $\mathbf{c} = \mathbf{R} \mathbf{R}^{\mathrm{T}} / T$ when all the entries $R_{ij}$ are drawn from a normal distribution $\mathcal{N}(0,\sigma)$. Quite importantly, this result is rigorously derived only in the thermodynamic limit \eqref{thermlim} of infinite matrix sizes growing to infinity at a fixed rate. In this limit, the Mar\v cenko-Pastur distribution reads

\begin{equation} \label{MP}
\rho_{\mathbf{c}} (\Lambda) = \frac{1}{2 \pi q \sigma^2} \frac{\sqrt{(\Lambda_+ - \Lambda) (\Lambda - \Lambda_-)}}{\Lambda} \ , \ \ \ \ \Lambda_{\pm} \doteq \sigma^2 \left ( 1 \pm \sqrt{q} \right )^2
\end{equation}
where $q$ is the rectangularity ratio defined in \eqref{rectratio}. However, as it can be seen in Figure \ref{SP_FTSE_spectra} (b)-(d), the Mar\v cenko-Pastur distribution actually provides a very poor fit of empirical distributions when $q$ and $\sigma$ are assumed to be equal to $N/T$ and 1 (for standardized data), respectively. The aforementioned eigenvalue bulks are reasonably well fitted by a Mar\v cenko-Pastur distribution only when $q$ and $\sigma$ are assumed to be \emph{free} parameters, whose values are to be determined via fitting. In particular, this typically causes $q$ to deviate from the ratio $N/T$, thus introducing the concept of effective system size. \\
Since the Mar\v cenko-Pastur distribution emerges as the limiting density for the covariance matrix of $N$ \emph{uncorrelated} time series made of $T$ observations, identifying eigenvalue bulks such as the ones in Figure \ref{SP_FTSE_spectra} with it basically amounts to state that most of the information contained in empirical covariance matrix spectra is actually no information at all, being equivalent to the spectrum one would obtain in the presence of pure noise \cite{Burda2004_1, Pafka}. On the other hand, this viewpoint allows one to give a specific meaning to the ``large'' eigenvalues out of the bulk. As it would also be possible to verify with Principal Component Analysis (PCA) \cite{Shlens}, such eigenvalues correspond to groups of correlated assets, most typically belonging to the same market sector. Analogously, the largest eigenvalue of the distribution is usually identified with the ``market mode'': such an eigenvalue appears as a consequence of those fluctuations that involve the market as a whole, and as a matter of fact the PCA can easily show it to account for a large part of the return variance. \\
As already anticipated, factor models \eqref{factormodel} represent good candidates to reproduce most of the empirical features shown in Figure \ref{SP_FTSE_spectra}. In the following sections, we shall make use of such models to challenge the previously mentioned common knowledge, according to which the eigenvalue bulks in empirical covariance matrix spectra essentially correspond to noise. Such a common knowledge has already been revised critically in a number of works (see for example \cite{Burda2004_2, Kwapien, Akemann, Guhr1, Guhr2}), and in this paper we wish to present an additional amount of evidence in this direction. \\
The paper is organized as follows. In Section II the ``direct'' problem of analytically estimating eigenvalue densities is addressed. In particular, some specific versions of the factor model in equation \eqref{factormodel} will be introduced and the eigenvalue spectra for the correlation matrix $\mathbf{C}$ of such model will be derived (sometimes performing approximations). Then, the RMT results provided in \cite{Burda2004_1, Burda2004_2} will be applied in order to derive exact results for the noise-dressed version $\mathbf{c}$ of the correlation matrix. Eventually, a subsection will be devoted to discuss the results obtained via Monte Carlo simulations in order to validate those analytical results. In the light of such numerical results, we shall also briefly discuss again the applicability limits of the Mar\v cenko-Pastur distribution. In Section III, the ``inverse'' problem of inferring eigenvalue densities from the empirically observed ones will be discussed. More specifically, a filtering procedure will be devised in order to highlight some of the cluster structure in empirical correlation matrices. Such a procedure will be performed on two data sets (relative to the S$\&$P500 and the FTSE350 Indices), and the results will be interpreted in terms of factor models. Eventually, in Section IV some conclusions and possible future perspectives of this work will be outlined.

\section{Theory: the direct problem}

\subsection{Cluster models: heuristic analysis}
	
Let us now specialize the factor model \eqref{factormodel}. In particular, let us start from the situation where all asset returns obey the following equation

\begin{equation} \label{commonfactor}
r_i(t) = \gamma_N m_N(t) + (1-\gamma_N) \epsilon_i(t),
\end{equation}
where $m_N(t), \epsilon_i(t) \sim \mathcal{N} (0,1) \ \forall t$ and $\gamma_N \in [0,1]$. In the previous equation $m_N$ represents a common mode driving all assets with the same ``intensity'' $\gamma_N$. We shall now build $K$ clusters of correlated assets from equation \eqref{commonfactor}. Thus, let there be $K$ groups of $N_k$ variables ($k = 1, \ldots, K$) with $\bar{N} \doteq \sum_{k=1}^K N_k \leq N$, and let us order the assets so that $r_i$ belongs to the $k$th assets for $i = 1+\sum_{l=1}^{k-1} N_l, \ldots, \sum_{l=1}^k N_l$. We can also denote the generic element in the $k$th cluster as $r_i^{(k)}$. We shall define it as

\begin{equation} \label{noclustereq}
r_i^{(k)}(t) = \gamma_k m_k(t) + (1-\gamma_k) r_i(t) \ , \ \ \ \ i = 1+\sum_{l=1}^{k-1} N_l, \ldots, \sum_{l=1}^k N_l
\end{equation}
where $\gamma_k \in [0,1]$, $m_k(t) \sim \mathcal{N}(0,1)$ is a cluster mode and $r_i$ is as in equation \eqref{commonfactor}. Thus, we can rewrite the previous relation as

\begin{equation} \label{clustereq}
r_i^{(k)}(t) = \gamma_k m_k(t) + (1-\gamma_k) \gamma_N m_N(t) + (1-\gamma_k)(1-\gamma_N) \epsilon_i(t) \ , \ \ \ \ i = 1+\sum_{l=1}^{k-1} N_l, \ldots, \sum_{l=1}^k N_l.
\end{equation}
We still simply call $r_i$ ($i = 1 + \bar{N}, \ldots, N$) those elements which do not belong to any cluster, and we assume them to evolve according to \eqref{commonfactor}. We always have $\mathbb{E}[r_i(t)] = \mathbb{E} \left [r_j^{(k)}(t) \right ] = 0 \ \forall, i,j,k,t$. Recalling the relations in \eqref{normalizations}, which can be generalized to include $m_N$ in a straightforward way, we can calculate all possible covariance matrix elements between assets described by \eqref{commonfactor} and \eqref{clustereq}. Four separate cases can be distinguished:

\begin{eqnarray} \label{covelements}
\mathbb{E}[r_i(t) r_j(t)] &=& (1-\gamma_N)^2 \delta_{ij} + \gamma_N^2 \\ \nonumber
\mathbb{E}\left [r_i(t)r_j^{(k)}(t) \right] &=& (1-\gamma_k)\gamma_N^2 \\ \nonumber
\mathbb{E}\left [r_i^{(k)}(t)r_j^{(k)}(t) \right] &=& (1-\gamma_k)^2 (1-\gamma_N)^2 \delta_{ij} + (1-\gamma_k)^2 \gamma_N^2 + \gamma_k^2 \\ \nonumber
\mathbb{E}\left [r_i^{(k)}(t)r_j^{(l)}(t) \right] &=& (1-\gamma_k) (1-\gamma_l)(1-\gamma_N)^2 \delta_{ij} + (1-\gamma_k)(1-\gamma_l)\gamma_N^2.
\end{eqnarray}
We are now in position to compute the correlation matrix $\mathbf{C}$ of the model, whose matrix elements read

\begin{equation}
C_{ij} = \frac{\mathbb{E}[r_i(t) r_j(t)]}{\sqrt{\mathrm{Var}[r_i(t)] \mathrm{Var}[r_j(t)]}}
\end{equation}
with straightforward generalizations to those involving elements belonging to clusters. \\
We shall focus for now on the limiting case in which correlations between cluster elements are very strong, \emph{i.e.} when $\gamma_k \rightarrow 1$ in each cluster. One can see from \eqref{covelements} that under this assumption the model's correlation matrix has a simple block-diagonal structure:

\begin{eqnarray} \label{covblock}
\mathbf{C} = \left ( \begin{array}{ccccc}
		\mathbf{E}^{(N_1)} & 0 & \ldots & 0 & 0 \\
		0 & \mathbf{E}^{(N_2)} & \ldots & 0 & 0 \\
		\vdots & \vdots & \ddots & \vdots & \vdots \\
		0 & 0 & \ldots & \mathbf{E}^{(N_K)} & 0 \\
		0 & 0 & \ldots & 0 & \mathbf{F}^{(N-\bar{N})} \\
	\end{array} \right).
\end{eqnarray}
where $\mathbf{E}^M$ is the $M \times M$ matrix whose entries are all equal to unity ($E^M_{ij} = 1 \ \forall i,j$), while $\mathbf{F}^{(N-\bar{N})}$ is a $(N-\bar{N}) \times (N-\bar{N})$ matrix with a slightly more complicated structure:

\begin{eqnarray}
\mathbf{F}^{(N-\bar{N})} = \left ( \begin{array}{ccccc}
		1\ & \frac{\gamma_N^2}{(1-\gamma_N)^2 + \gamma_N^2} & \ldots & \frac{\gamma_N^2}{(1-				\gamma_N)^2 + \gamma_N^2} \\
		\frac{\gamma_N^2}{(1-\gamma_N)^2 + \gamma_N^2} & 1 & \ldots & \frac{\gamma_N^2}{(1-				\gamma_N)^2 + \gamma_N^2} \\
		\vdots & \vdots & \ddots & \vdots \\
		\frac{\gamma_N^2}{(1-\gamma_N)^2 + \gamma_N^2} & \frac{\gamma_N^2}{(1-\gamma_N)^2 + 			\gamma_N^2} & \ldots & 1 \\
	\end{array} \right).
\end{eqnarray}
The block structure in \eqref{covblock} allows for the computation of the eigenvalue spectrum. In fact, since we have

\begin{eqnarray}
\mathrm{det} \left (\mathbf{E}^{(M)} - \Lambda \mathbf{I}_M \right) &=& \Lambda^{M-1} (M - \Lambda) \\ \nonumber
\mathrm{det} \left (\mathbf{F}^{(N-\bar{N})} - \Lambda \mathbf{I}_{N-\bar{N}} \right ) &=& \left ( \frac{(N-\bar{N}) \gamma_N^2 + (1-\gamma_N)^2}{(1-\gamma_N)^2 + \gamma_N^2} - \Lambda \right ) \left ( \frac{(1-\gamma_N)^2}{(1-\gamma_N)^2 + \gamma_N^2} - \Lambda \right )^{N-\bar{N}-1}
\end{eqnarray} 
the characteristic equation for the $\mathbf{C}$ matrix reads:

\begin{equation} \label{thspectrum}
\Lambda^{\bar{N}-K} \left ( \Lambda - \frac{(N-\bar{N}) \gamma_N^2 + (1-\gamma_N)^2}{(1-\gamma_N)^2 + \gamma_N^2} \right ) \left ( \Lambda - \frac{(1-\gamma_N)^2}{(1-\gamma_N)^2 + \gamma_N^2} \right )^{N-\bar{N}-1} \prod_{k=1}^K (\Lambda - N_k) = 0.
\end{equation}
This eigenvalue spectrum is able to reproduce, at least on a heuristic level, some of the features of empirical spectra (see Figure \ref{SP_FTSE_spectra}): each cluster gives rise to a large eigenvalue equal to the cluster size $N_k$, and the common mode produces one large eigenvalue $\sim (N-\bar{N})\gamma_N^2 / ((1-\gamma_N)^2 + \gamma_N^2)$ too. It is worth mentioning that this latter eigenvalue might not necessarily be the largest one: as a matter of fact, large enough $N_k$s and a small $\gamma_N$ can lead to situations in which the largest eigenvalue is given by $\mathrm{max}_{k} \ N_k$. Even though this seems not to be the case in most financial applications, it is still worth stressing that the largest eigenvalue in empirical spectra should not be labelled as the ``market eigenvalue'' right away, but only after some further checks (as, for example, the inspection of the corresponding eigenvector). \\
Going back to equation \eqref{thspectrum}, a $(N-\bar{N}-1)$-fold degenerate eigenvalue (equal to $(1-\gamma_N)^2 / ((1-\gamma_N)^2 + \gamma_N^2)$) can be recognized. Also, equation \eqref{thspectrum} indicates that each cluster gives rise to $N_k - 1$ zero modes, altogether forming a group of $\bar{N} - K$ zero modes. In a noise-marred situation, as it can be verified by means of Monte Carlo simulations, the degeneracies in \eqref{thspectrum} are broken and give rise to two bulks. In the highly correlated cluster assumption ($\gamma_k \rightarrow 1$) yielding \eqref{thspectrum} the two bulks typically remain well separated. However, when such assumption is relaxed, allowing for small values of the $\gamma_k$s, the two bulks get closer, and for properly chosen values of the parameters they eventually ``collide'' and merge into one single structure (see Subsection II C and the figures in it). This latter might be identified with the typical eigenvalue bulks appearing in empirical spectra (see Figure \ref{SP_FTSE_spectra}). It is important to stress, already at this heuristic level, that the emergence of such a bulk in this factor model stems from the presence of (weak) correlations between the assets, oppositely to the Mar\v cenko-Pastur distribution \eqref{MP}, which in turn, as already discussed, originates from pure noise. Nevertheless, quite subtly the Mar\v cenko-Pastur distribution can still provide good fits to such bulks in a number of situations, as we shall illustrate later. \\
The previous factor model, yielding equation \eqref{thspectrum} for the eigenvalue spectrum of its correlation matrix, can be further simplified to the case where no common factor is driving the asset returns. This can be directly achieved on the eigenvalue spectrum by setting $\gamma_N = 0$ in equation \eqref{thspectrum}. This gives

\begin{equation} \label{thspectrum2}
\Lambda^{\bar{N}-K} (\Lambda -1)^{N-\bar{N}} \prod_{k=1}^K (\Lambda - N_k) = 0.
\end{equation}
This spectrum still yields one large eigenvalue for each cluster and two degenerate eigenvalues equal to zero and one, respectively. Just like in the previous discussion, let us now relax the assumption of strong correlations ($\gamma_k \rightarrow 1$) within clusters. For the sake of simplicity, let us assume that all assets in each cluster are mutually correlated with the same correlation coefficient $\rho_k \in [0,1]$ (which can be explicitly computed from $\eqref{covelements}$). So, the correlation matrix of the model would read

\begin{eqnarray} \label{covblock2}
\mathbf{C} = \left ( \begin{array}{ccccc}
		\tilde{\mathbf{E}}^{(N_1)} & 0 & \ldots & 0 & 0 \\
		0 & \tilde{\mathbf{E}}^{(N_2)} & \ldots & 0 & 0 \\
		\vdots & \vdots & \ddots & \vdots & \vdots \\
		0 & 0 & \ldots & \tilde{\mathbf{E}}^{(N_K)} & 0 \\
		0 & 0 & \ldots & 0 & \mathbf{I}_{N-\bar{N}} \\
	\end{array} \right)
\end{eqnarray}
where each $\tilde{\mathbf{E}}^{(N_k)}$ is a $N_k \times N_k$ matrix such that $\tilde{E}^{(N_k)}_{ij} = \rho_k$ for $i \neq j$ and $\tilde{E}^{(N_k)}_{ii} = 1$, while the last block is now given by the identity matrix (as it can be seen from the first relation in \eqref{covelements} for $\gamma_N = 0$). One can verify that

\begin{equation} \label{seceq}
\mathrm{det} \left ( \tilde{\mathbf{E}}^{(N_k)} - \Lambda \mathbf{I}_{N_k} \right ) = \left [\Lambda - (1 - \rho_k) \right ]^{N_k -1 } \left [\Lambda - (N_k \rho_k + (1 - \rho_k)) \right ].
\end{equation}
In order to further simplify things, let us consider the case where we have just one cluster of $\bar{N}$ assets with mutual correlation $\rho$. Equation \eqref{thspectrum} in this case would need to be modified to read

\begin{equation} \label{thspectrum3}
\left [\Lambda - (1 - \rho) \right ]^{\bar{N}-1} (\Lambda -1)^{N-\bar{N}} [\Lambda - (\bar{N} \rho + (1-\rho))] = 0,
\end{equation}
thus giving one large eigenvalue, and two degenerate eigenvalues equal to $(1-\rho)$ and one. The latter emerges as a consequence of the $N-\bar{N}$ mutually uncorrelated assets, \emph{i.e.} as a consequence of pure noise, while the former is due to the presence of a cluster. Just like in the case discussed previously, a noise-dressed version of \eqref{thspectrum3} would lead to two eigenvalue bulks, and suitably chosen values of $\rho$ would make the two bulks merge into one (see Subsection II C). Thus, in this case too, the emergence of a main bulk would not be a consequence of pure noise alone. \\

\subsection{Cluster models: exact results}

The proper mathematical framework to deal with covariance matrices featuring degenerate spectra (as the ones in equations \eqref{thspectrum}, \eqref{thspectrum2} and \eqref{thspectrum3}) is the one provided in \cite{Burda2004_1, Burda2004_2}, and we shall exploit it extensively in the following. So, first let us introduce some basic notions and notations of RMT. Just like we did so far, we shall denote the eigenvalues of the correlation matrix $\mathbf{C}$ of a given model as $\Lambda_i$ ($i = 1, \ldots, N$), while the eigenvalues of the corresponding estimator \eqref{Pearson} will be denoted as $\lambda_i$. Quite straightforwardly, one can define the eigenvalue density for the theoretical correlation matrix as

\begin{equation} \label{thdensity}
\rho_{\mathbf{C}}(\Lambda) = \frac{1}{N} \sum_{i=1}^N \delta (\Lambda - \Lambda_i),
\end{equation}
and this is related to the matrix moments $M_{\mathbf{C}}^{(k)}$:

\begin{equation} \label{thmoments}
M_{\mathbf{C}}^{(k)} \doteq \frac{1}{N} \mathrm{Tr}\ \mathbf{C}^k = \frac{1}{N} \sum_{i=1}^N \Lambda_i^k =
\int \mathrm{d}\Lambda \rho_{\mathbf{C}}(\Lambda) \Lambda^k.
\end{equation}
In analogy to \eqref{thdensity}, one can define an expected spectral density for the estimator $\mathbf{c}$ in equation \eqref{Pearson}:

\begin{equation} \label{estdensity}
\rho_{\mathbf{c}}(\lambda) = \frac{1}{N} \sum_{i=1}^N \mathbb{E} \left [ \delta (\lambda - \lambda_i) \right ],
\end{equation}
where the expectation is to be meant with respect to the probability measure \eqref{probmeas}. Generalizing \eqref{thmoments}, we can then define the expected matrix moments as

\begin{equation} \label{estmoments}
m_{\mathbf{c}}^{(k)} \doteq \frac{1}{N} \mathbb{E} \left [ \mathrm{Tr} \mathbf{c}^k \right ] = \int \mathrm{d} \lambda \rho_{\mathbf{c}} (\lambda) \lambda^k.
\end{equation}
The two corresponding resolvents, or Green's functions, are given by:

\begin{eqnarray} \label{Green}
\mathbf{G}_{\mathbf{C}}(Z) &=& \left ( Z \mathbf{I}_N - \mathbf{C} \right )^{-1} \\ \nonumber
\mathbf{g}_{\mathbf{c}}(z) &=& \mathbb{E} \left [ \left ( z \mathbf{I}_N - \mathbf{c} \right )^{-1} \right ]
\end{eqnarray}
where $Z, z \in \mathbb{C}$. Then, one can introduce the moment generating functions, and it is possible to show that they are closely related to the Green's functions in the following way

\begin{eqnarray} \label{mgf}
M_{\mathbf{C}}(Z) &\doteq& \sum_{k=1}^{\infty} \frac{M_{\mathbf{C}}^{(k)}}{Z^k} = Z G_{\mathbf{C}}(Z) - 1 \\ \nonumber
m_{\mathbf{c}}(z) &\doteq&  \sum_{k=1}^{\infty} \frac{m_{\mathbf{c}}^{(k)}}{z^k} = z g_{\mathbf{c}}(z) - 1,
\end{eqnarray}
where we have 

\begin{eqnarray} \label{normtrace}
G_{\mathbf{C}}(Z) &\doteq& \frac{\mathrm{Tr} [\mathbf{G}_{\mathbf{C}}(Z)]}{N} \\ \nonumber
g_{\mathbf{c}}(z) &\doteq& \frac{\mathrm{Tr} [\mathbf{g}_{\mathbf{c}}(z)]}{N}. 
\end{eqnarray}
Moreover, from the well known relation $\lim_{\epsilon \rightarrow 0^+} (\lambda + \mathrm{i} \epsilon)^{-1} = \mathcal{P} (\lambda^{-1}) - \mathrm{i} \pi \delta(\lambda)$ (where $\mathcal{P}$ denotes the principal value), one can show that the eigenvalue densities \eqref{thdensity} and \eqref{estdensity} can be directly derived from the corresponding Green's functions \eqref{Green}:

\begin{eqnarray} \label{rhofromG}
\rho_{\mathbf{C}} (\Lambda) &=& - \frac{1}{\pi} \lim_{\epsilon \rightarrow 0^+} \mathrm{Im} \ G_{\mathbf{C}} (\Lambda + \mathrm{i} \epsilon) \\ \nonumber
\rho_{\mathbf{c}} (\lambda) &=& - \frac{1}{\pi} \lim_{\epsilon \rightarrow 0^+} \mathrm{Im} \ g_{\mathbf{c}} (\lambda + \mathrm{i} \epsilon).
\end{eqnarray} 
So, basically, the Green's function contains the same information as the whole eigenvalue density, and the same, through \eqref{mgf}, is also true for the moment generating function. In particular, for $\Lambda, \lambda > 0$, the previous relations can be converted into: \\

\begin{eqnarray} \label{rhofromM}
\rho_{\mathbf{C}} (\Lambda) &=& - \frac{1}{\pi \Lambda} \lim_{\epsilon \rightarrow 0^+} \mathrm{Im} \ M_{\mathbf{C}} (\Lambda + \mathrm{i} \epsilon) \\ \nonumber
\rho_{\mathbf{c}} (\lambda) &=& - \frac{1}{\pi \lambda} \lim_{\epsilon \rightarrow 0^+} \mathrm{Im} \ m_{\mathbf{c}} (\lambda + \mathrm{i} \epsilon).
\end{eqnarray}
A fundamental relation between between the moment generating functions of a ``true'' correlation matrix and its estimator in the infinite matrix size limit \eqref{thermlim} can be derived \cite{Burda2004_1, Burda2004_2} either in the framework of Free Random Variables \cite{Burda2010, Barndorff-Nielsen, Politi} or using planar diagrammatic methods \cite{Burda2004_1, Itzykson}. This derivation will be outlined in Appendix A. The starting point is the following simple relation between moment generating functions

\begin{equation} \label{mMrel}
m_{\mathbf{c}}(z) = M_{\mathbf{C}}(Z),
\end{equation}
where the two complex arguments are related by the following transformation:

\begin{equation} \label{confmap}
Z = \frac{z}{1 + q m_{\mathbf{c}}(z)}.
\end{equation}
Once $M_{\mathbf{C}}(Z)$ is known, $m_{\mathbf{c}}(z)$ can be derived in principle from the following functional equation:

\begin{equation} \label{funcrel}
m_{\mathbf{c}}(z) = M_{\mathbf{C}} \left ( \frac{z}{1 + q m_{\mathbf{c}}} \right ).
\end{equation}

Bearing in mind the previous discussion on factor models, we shall focus on correlation matrices whose spectra display degenerate eigenvalues. Let us then assume the correlation matrix $\mathbf{C}$ to have $L$ distinct eigenvalues $\Lambda_i$ ($i = 1, \ldots, L$) with degeneracies $n_i$. The moment generating function for such a matrix is given by

\begin{equation} \label{mgfC}
M_{\mathbf{C}} (Z) = \frac{1}{N} \sum_{i=1}^L \frac{n_i \Lambda_i}{Z - \Lambda_i} = \sum_{i=1}^L \frac{w_i \Lambda_i}{Z - \Lambda_i}
\end{equation}
where the weights $w_i = n_i / N$ have been introduced. Thus, from \eqref{funcrel} we get

\begin{equation} \label{meq}
m_{\mathbf{c}}(z) = \sum_{i=1}^L \frac{w_i \Lambda_i (1 + q m_{\mathbf{c}}(z))}{z - \Lambda_i (1 + q m_{\mathbf{c}}(z))}.
\end{equation}
For each fixed $z$, this becomes a polynomial equation of degree $L+1$ in $m_{\mathbf{c}}(z)$m, yielding as many solutions. The problem arises of choosing the right one: as extensively discussed and detailed in \cite{Burda2006}, the right branch of the map in equation \eqref{confmap} to pick up is the one giving $Z \rightarrow z$ for $z \rightarrow \infty$. In the simplest case one has $\mathbf{C} = \mathbf{I}_N$ and, of course, the correlation matrix has just one $N$-fold degenerate eigenvalue equal to one: it can be shown that, in this case, equation \eqref{meq} leads precisely to the Mar\v cenko-Pastur distribution \eqref{MP}, as one would expect. On the other hand, already when considering two distinct eigenvalues, quite different scenarios are possible, including the previously discussed cases of well separated or merging bulks (see the next subsection). Let us also remark that equation \eqref{meq} cannot be applied to the large non-degenerate eigenvalues typically displayed by factor models. This is because, as already stated, we shall always work in the thermodynamic limit \eqref{thermlim}, where the weight ($1/N$) of such eigenvalues vanishes as $N \rightarrow \infty$. As a matter of fact, this kind of eigenvalues need to be investigated \emph{per se}, and actually extensive areas of the RMT literature are devoted to the study of statistical properties of single eigenvalues as well as order statistics \cite{Tracy}. In particular, it has been shown in \cite{Paul} that large non-degenerate sample eigenvalues of correlation matrices follow a normal distribution (see the next subsection for a numerical confirmation).

\subsection{Monte Carlo simulations}
\begin{figure}
\begin{center}
\includegraphics[width=2.9 in, height=2.0 in]{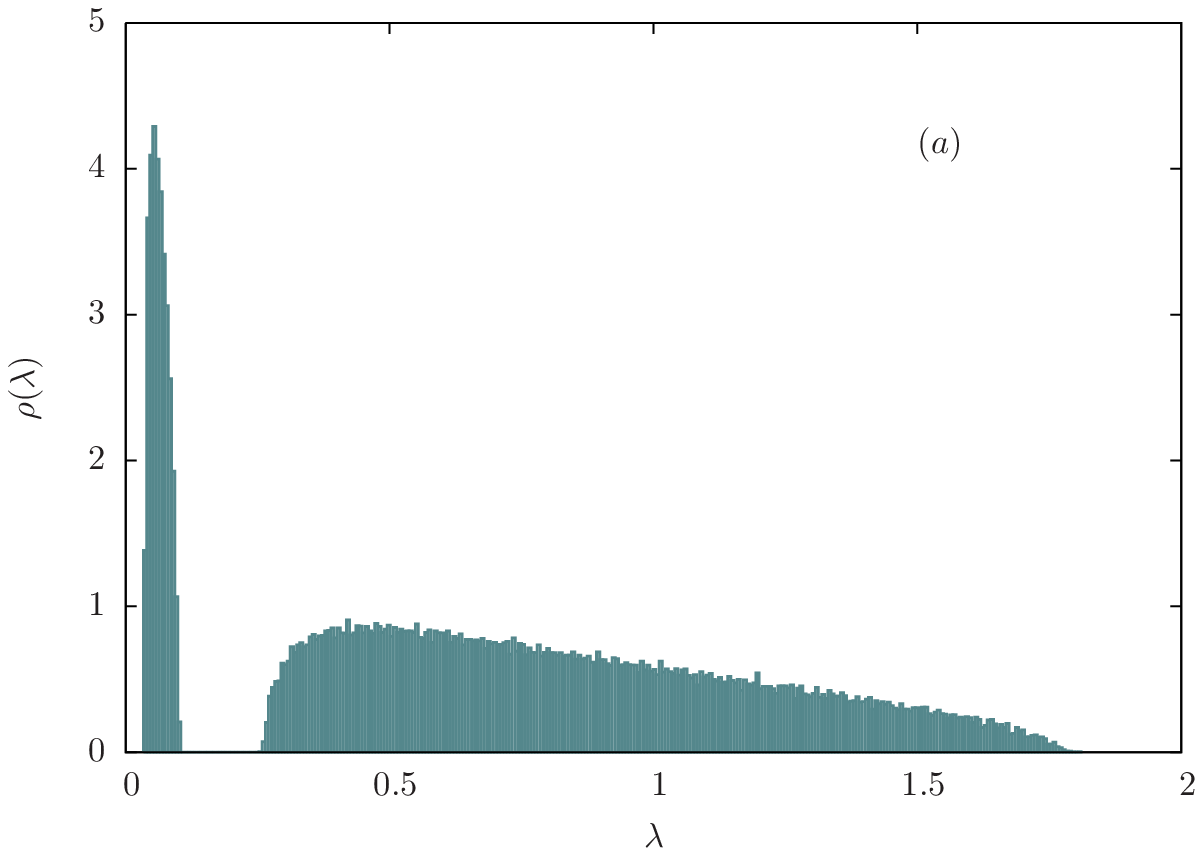} 
\hspace{0.1 in} 
\includegraphics[width=2.9 in, height= 2.0 in]{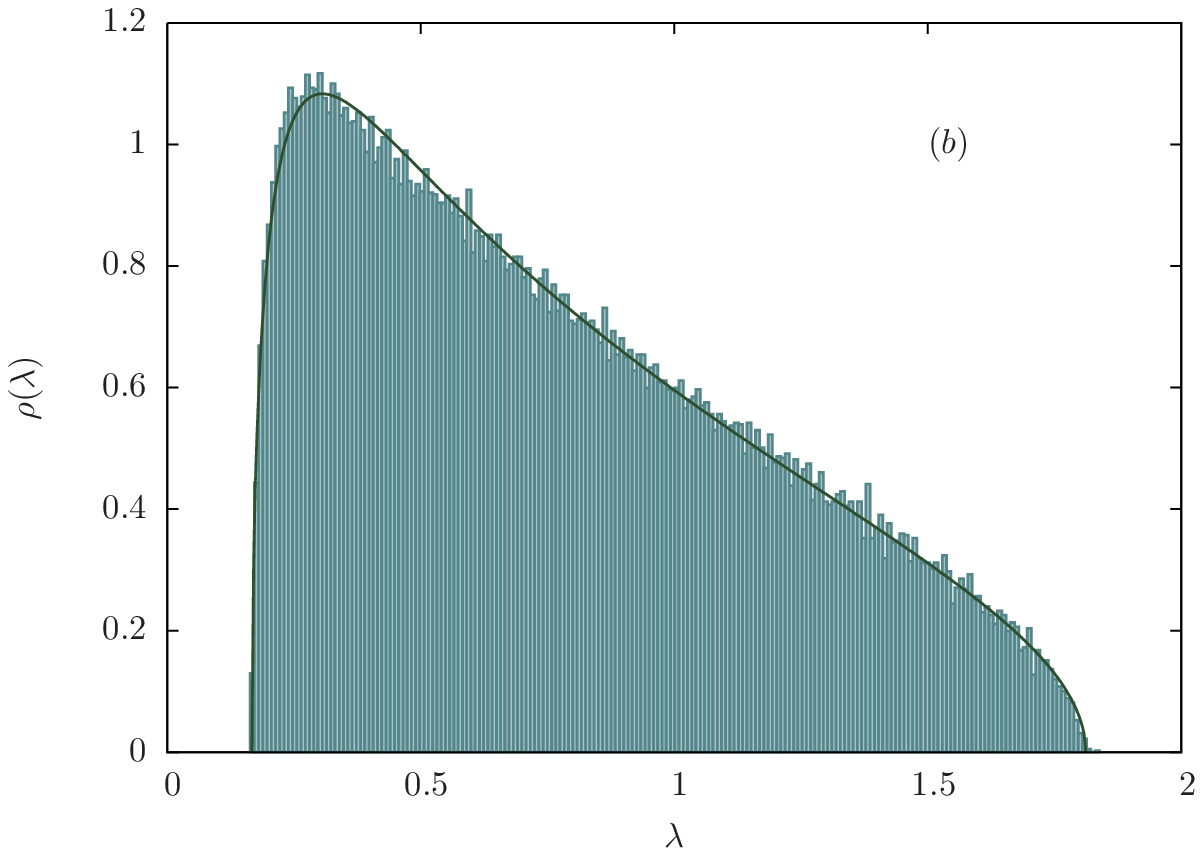} 
\caption{(a) Eigenvalue density for $100$ simulations of the factor model described in \eqref{noclustereq}, \eqref{clustereq} and \eqref{covelements} with $N = 500$, $T = 2000$ ($q = 0.25$). One cluster made of $N_1 = \bar{N} = 100$ variables, correlated via a coefficient $\gamma_1 = 0.7$, is present. A common mode is introduced via a coefficient $\gamma_N = 0.3$. As expected from equation \eqref{thspectrum}, two eigenvalue bulks are clearly visible. The mean value in the bulk on the left is $0.04$, while equation \eqref{thspectrum} would predict zero as a consequence of the strong correlation limit ($\gamma_k \rightarrow 1$) approximation. On the other hand, the mean value in the bulk on the right is $0.85$, in remarkable agreement with the predicted value $(1-\gamma_N)^2 / ((1-\gamma_N)^2 + \gamma_N^2) = 0.84$. For the sake of readability, the ``large'' eigenvalues are not shown. (b) By setting $\gamma_1 = 0.4$, the two bulks in (a) merge into a single one. Such a structure, despite emerging as a consequence of (weak) correlations, is very well fitted by a Mar\v cenko-Pastur distribution (see equation \eqref{MP}) with $q=0.29$ and $\sigma=0.88$.} \label{Factor_models_spectra1}
\end{center}
\end{figure}

\begin{figure}
\begin{center}
\includegraphics[width=2.9 in, height=2.0 in]{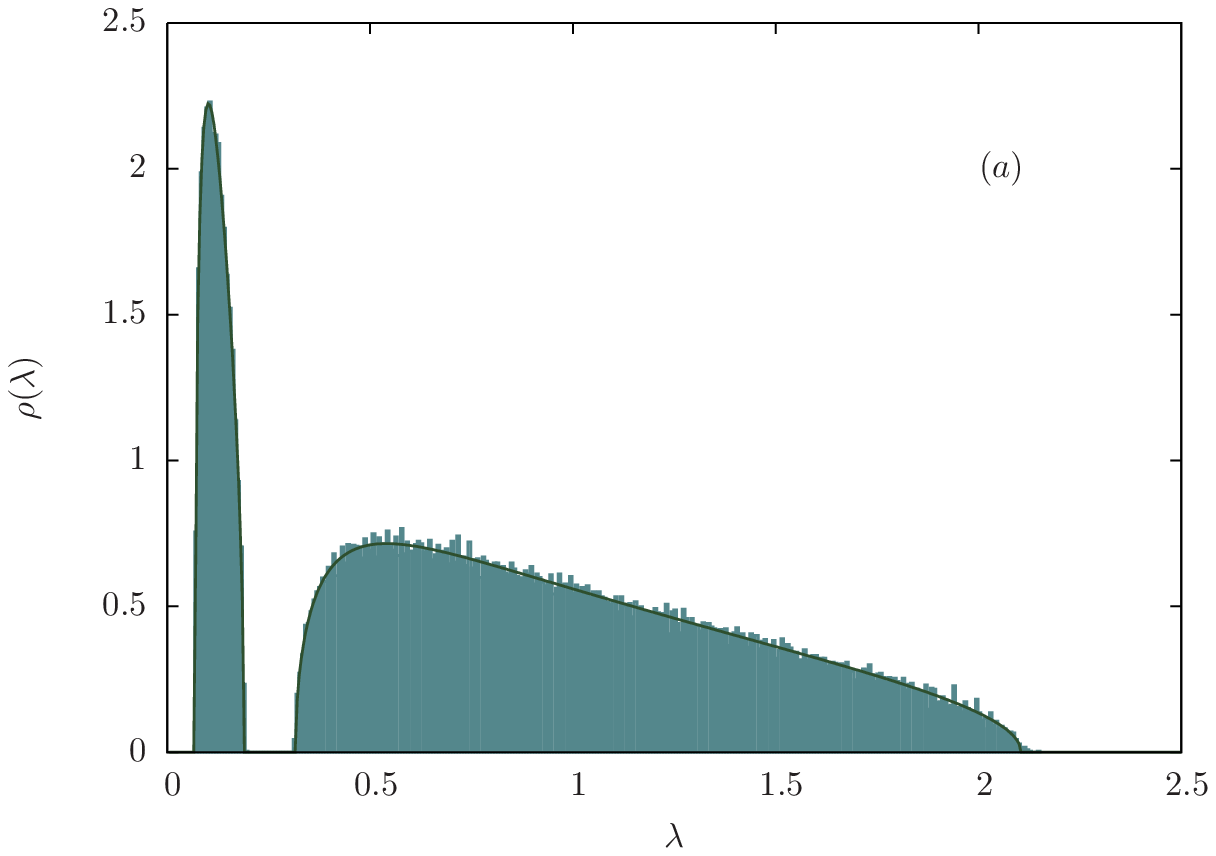} 
\hspace{0.1 in} 
\includegraphics[width=2.9 in, height= 2.0 in]{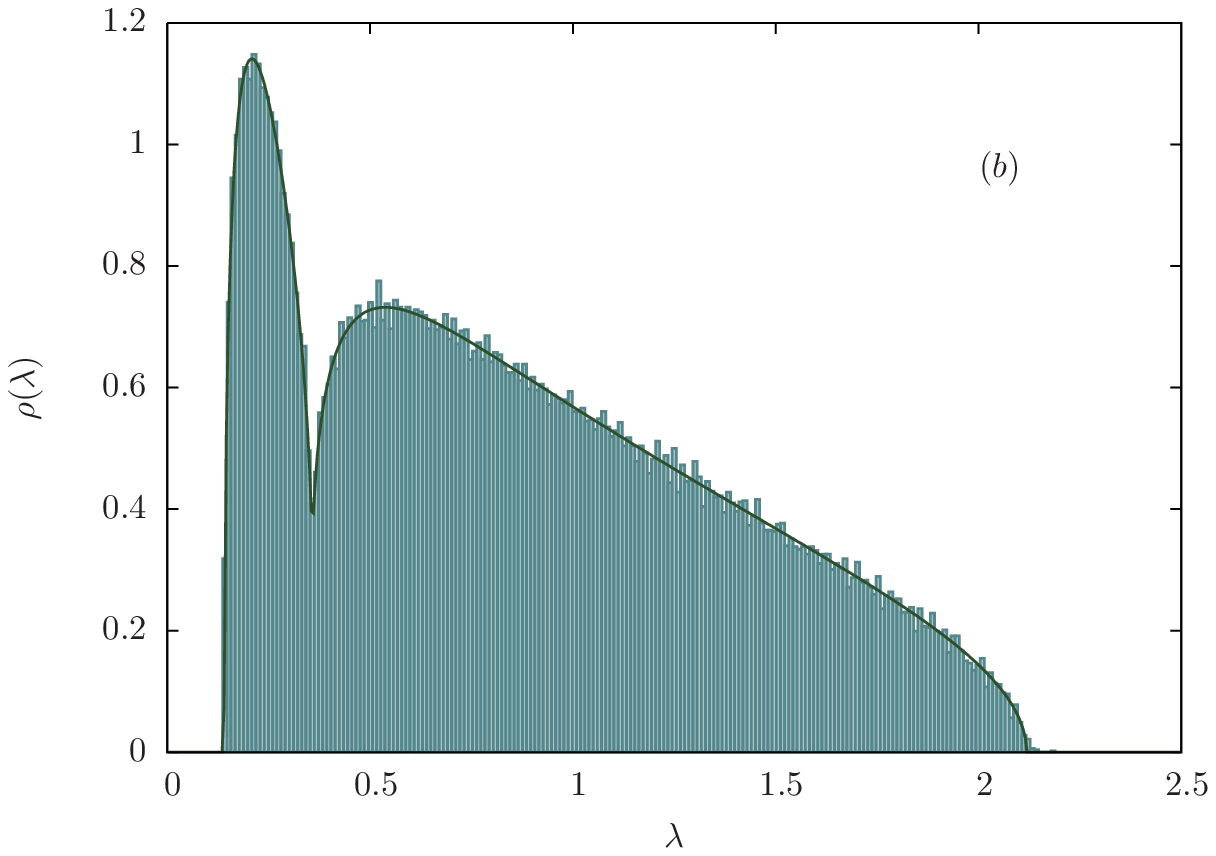} 
\vspace{0.1 in}
\includegraphics[width=2.9 in, height= 2.0 in]{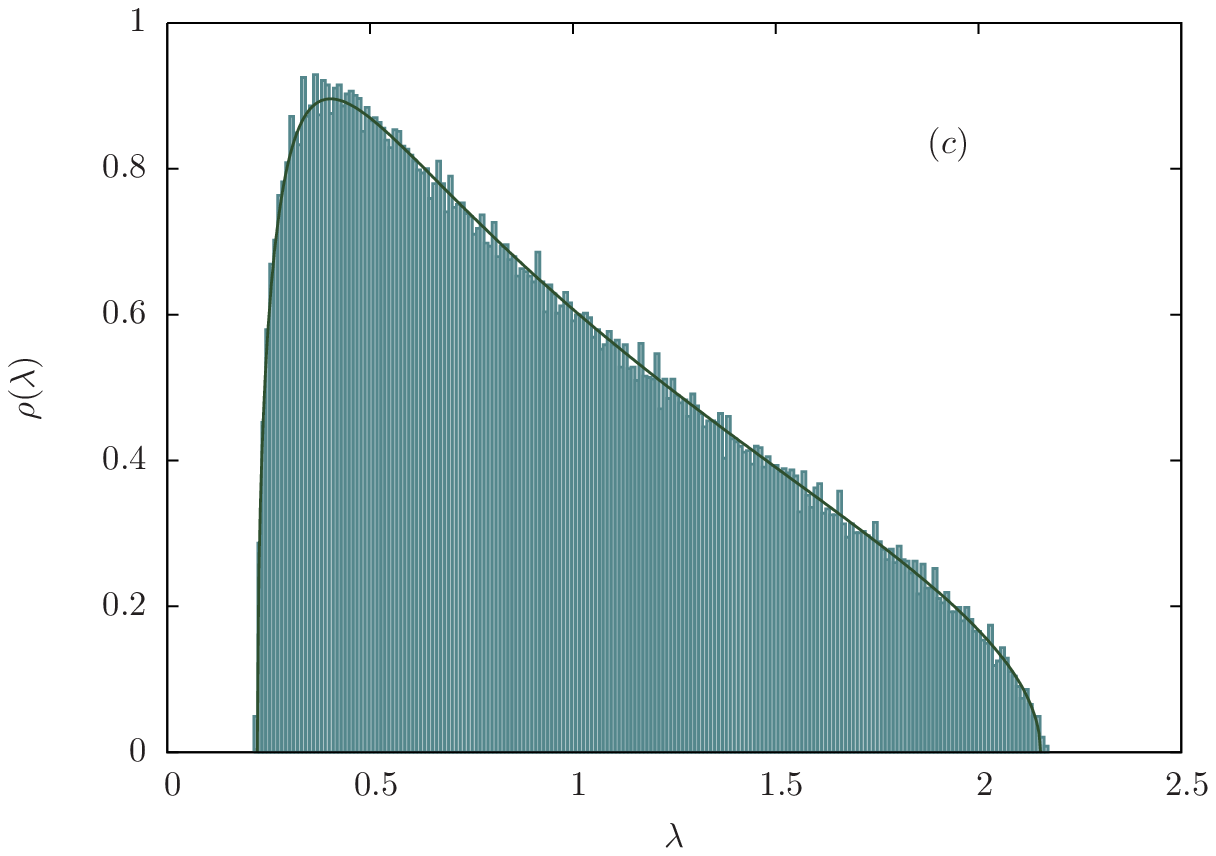}
\hspace{0.1 in} 
\includegraphics[width=2.9 in, height= 2.0 in]{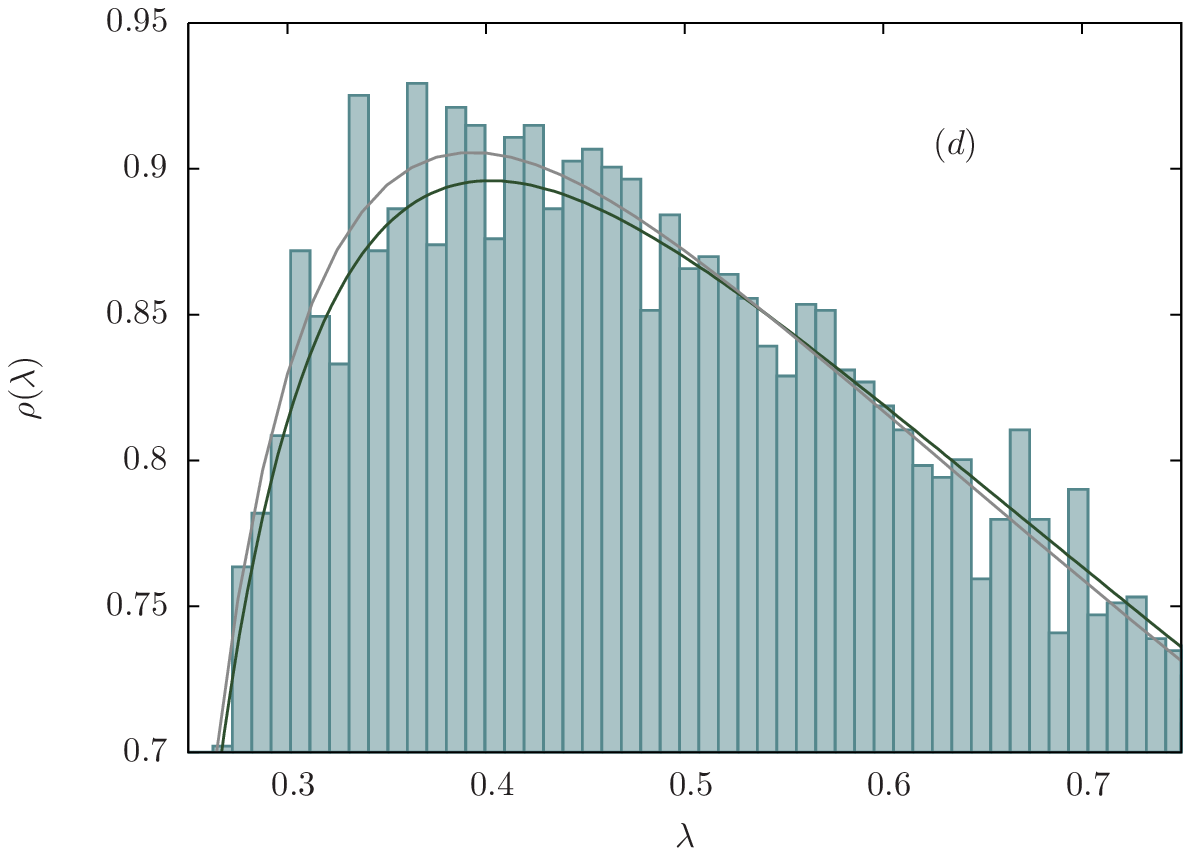} 
\caption{(a) Eigenvalue spectrum of the correlation matrix for the factor model yielding the spectrum in \eqref{thspectrum3} for $N=500$, $T=2000$, $\bar{N}=100$ and $\rho = 0.84$. This model yields two degenerate eigenvalues: $\Lambda_1 = 1-\rho = 0.16$ and $\Lambda_2 = 1$ (see also Figure \ref{Factor_models_spectra1}). The histogram is the result of $100$ Monte Carlo simulations of such model, while the solid line represents the density obtained from the solution of equation \eqref{meq}. (b) Eigenvalue spectrum for the same model with $\rho = 0.65$, \emph{i.e.} for $\Lambda_1 = 0.35$. It can be clearly seen that the two separated bulks shown in (a) start to merge as a consequence of the smaller correlations (smaller value of $\rho$). (c) Posing $\rho = 0.30$ the two eigenvalue bulks merge completely into one single structure. In analogy to Figure \ref{Factor_models_spectra1} (b), such a structure is apparently well fitted by a Mar\v cenko-Pastur distribution with $q=0.26$ and $\sigma=0.97$, plotted as a solid line. On this scale, the Mar\v cenko distributions would be barely distinguishable from the density obtained from equation \eqref{meq} with $\Lambda_1 = 1- \rho = 0.7$ and $\Lambda_2 = 1$. (d) Comparison between two such densities in correspondence of their peak, where they differ the most. Despite the quite small deviation between the two, a Kolmogorov-Smirnov (KS) performed on the data gave the following results. The critical values, for different significance levels $\alpha$, are given by $\mathrm{CV}_{\mathrm{KS}}(\alpha = 0.10) = 5.5 \times 10^{-3}$, $\mathrm{CV}_{\mathrm{KS}}(\alpha = 0.05) = 6.1 \times 10^{-3}$ and $\mathrm{CV}_{\mathrm{KS}}(\alpha = 0.01) = 7.3 \times 10^{-3}$. Under a null hypothesis of data distributed according to the Mar\v cenko-Pastur distribution, the value of the KS statistic was $\mathrm{STAT}_{\mathrm{KS}} = 7.9 \times 10^{-3}$, allowing for the rejection of the null hypothesis for all the significance levels considered. On the other hand, under the null assumption of data distributed according to the density obtained from equation \eqref{meq}, we obtained $\mathrm{STAT}_{\mathrm{KS}} = 2.3 \times 10^{-3}$, thus preventing from rejecting the null hypothesis. Clearly, the large statistics in this example plays a relevant role in helping the KS test to ``distinguish'' the two densities. Smaller data samples would prevent the Mar\v cenko-Pastur from being rejected.} \label{Factor_models_spectra2}
\end{center}
\end{figure}

\begin{figure}
\begin{center}
\includegraphics[width=2.9 in, height=2.0 in]{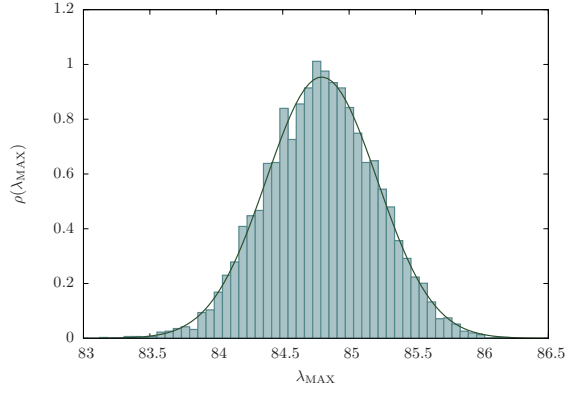} 
\hspace{0.1 in} 
\caption{Distribution of the largest eigenvalue $\lambda_{\mathrm{MAX}}$ from $5000$ Monte Carlo simulations of the factor model yielding the spectrum in \eqref{thspectrum3} with $\rho = 0.85$, $N = 500$ and $\bar{N} = 100$. The distribution is well fitted by a Normal distribution with expected value $m = 84.79$, very close to the theoretical value predicted by equation \eqref{thspectrum3}: $\bar{N} \rho + (1-\rho) = 85.15$. Three different statistical tests (Jarque-Bera, Lilliefors and Kolmogorov-Smirnov) \cite{Stephens} were performed, assuming a null hypothesis of Normally distributed data. In the following we report the different critical values (CV) obtained for the different tests and for different significance levels $\alpha$. Also, we report the statistic values (STAT), which, if smaller than the critical values, prevent the null hypothesis from being rejected. Jarque-Bera test: $\mathrm{STAT}_{\mathrm{JB}} = 3.114$, $\mathrm{CV}_{\mathrm{JB}}(\alpha = 0.10) = 4.605$, $\mathrm{CV}_{\mathrm{JB}}(\alpha = 0.05) = 5.992$, $\mathrm{CV}_{\mathrm{JB}}(\alpha = 0.01) = 9.210$. Lilliefors test: $\mathrm{STAT}_{\mathrm{L}} = 0.78 \times 10^{-2}$, $\mathrm{CV}_{\mathrm{L}}(\alpha = 0.10) = 1.14 \times 10^{-2}$, $\mathrm{CV}_{\mathrm{L}}(\alpha = 0.05) = 1.25 \times 10^{-2}$ and $\mathrm{CV}_{\mathrm{L}}(\alpha = 0.01) = 1.56 \times 10^{-2}$. Kolmogorov-Smirnov test: $\mathrm{STAT}_{\mathrm{KS}} = 0.78 \times 10^{-2}$, $\mathrm{CV}_{\mathrm{KS}}(\alpha = 0.10) = 1.73 \times 10^{-2}$, $\mathrm{CV}_{\mathrm{KS}}(\alpha = 0.05) = 1.92 \times 10^{-2}$, $\mathrm{CV}_{\mathrm{KS}}(\alpha = 0.01) = 2.30 \times 10^{-2}$. All of the previous results prevent from rejecting the hypothesis of Normally distributed data.} \label{Largest_eig}
\end{center}
\end{figure}

In this subsection we present and detail the Monte Carlo simulations we performed in order to test and validate the analytical results described so far. In all cases, we generated $T$ realizations of $N$ stochastic processes described by the factor model introduced in equations \eqref{noclustereq}, \eqref{clustereq} and \eqref{covelements} (from a numerical viewpoint, this just boils down to the generation of standard Gaussian random numbers). By choosing different parameter values, we implemented the different versions of the model which were discussed in the previous subsection, corresponding to different theoretical correlation matrices (equation \eqref{covblock}, \eqref{covblock2}).  The eigenvalues of the corresponding estimators \eqref{Pearson} were obtained via numerical diagonalization (by means the diagonalization algorithm provided by Matlab\textsuperscript{\textregistered}). \\
In Figure \ref{Factor_models_spectra1}, a first example of eigenvalue spectra deriving from factor models is presented.
In this first example a common mode (introduced via a non-zero $\gamma_N$ coefficient, see the figure caption for all the details on parameter values) as well as a correlated cluster of variables are present. As already discussed in the previous subsection, the degeneracies in equation \eqref{thspectrum} are broken, and, in the limit of strong correlations in the cluster ($\gamma_k \rightarrow 1$), two well separated eigenvalue bulks emerge (Figure \ref{Factor_models_spectra1} (a)). On the other hand, when such correlations get weaker, the two eigenvalue bulks get closer, eventually melting into one single structure (Figure \ref{Factor_models_spectra1}). Remarkably, such a structure is quite well fitted by a Mar\v cenko-Pastur distribution, which is however characterized by values of the $q$ and $\sigma$ parameters that differ  from the ones which would be obtained for standardized uncorrelated data ($q = N/T$ and $\sigma = 1$). \\
Such features are further illustrated in Figure \ref{Factor_models_spectra2}, which refers to the case of a factor model with no common mode ($\gamma_N = 0$). 
Again, the progressive fusion (induced by weaker correlations) between separated eigenvalue bulks is shown. Also, we compare the numerically obtained spectra to the eigenvalue densities obtained from the solution of equation \eqref{meq}, obtaining a very good agreement between the two (Figure \ref{Factor_models_spectra2} (a)-(b)). Just like in the previous case, the Mar\v cenko-Pastur distribution seems to provide quite a good fit of the ``limiting'' eigenvalue bulk obtained for small correlations (Figure \ref{Factor_models_spectra2} (c)). However, we performed a Kolmogorov-Smirnov \cite{Stephens} test under the null hypothesis of data distributed according to a Mar\v cenko-Pastur distribution, and we found such hypothesis to be rejected for all the significance levels we considered (see the caption of Figure \ref{Factor_models_spectra2} for further details). On the other hand, the same test prevented us from rejecting the hypothesis of data distributed according to the eigenvalue density obtained from the solution of equation \eqref{meq}, its degenerate eigenvalues being given by equation \eqref{thspectrum3}. This is quite surprising, given the great similarity between the two densities (see Figure \ref{Factor_models_spectra2} (d)), which would be almost undistinguishable if plotted on the scale of the whole distribution (as in Figure \ref{Factor_models_spectra2} (c)). Nevertheless, we believe this result to be quite relevant, since it strongly suggests that the theoretical framework outlined in the previous subsection might be the right way to describe and analyze empirical correlation matrices which display a cluster correlation structure, at least to some extent. In the following section, we shall apply these ideas to financial data. \\
We also believe these findings to provide some interesting evidence against the use of the Mar\v cenko-Pastur distribution whenever non-negligible correlations are present between random variables. Despite being close, in a number of situations, to the eigenvalue densities deriving from the solution of equation \eqref{meq}, the Mar\v cenko-Pastur distribution always needs to be fitted on the data under study, even when they are completely under control (as in the case of Monte Carlo simulations). Then, as already pointed out, the presence of correlations causes the parameters $q$ and $\sigma$ to deviate from the corresponding values which would be obtained in a pure noise situation. In particular, given the definition in equation \eqref{rectratio}, this leads to the introduction of the artificial, and possibly misleading, concept of effective system size. \\
Eventually, concluding this subsection on Monte Carlo simulations, in Figure \ref{Largest_eig} we show a numerically obtained distribution of the largest sample eigenvalue for a factor model yielding the eigenvalue spectrum in \eqref{thspectrum3}. As it can be seen by direct inspection, the corresponding histogram is well fitted by a Gaussian distribution, as already anticipated in the previous subsection. Moreover, three statistical tests (whose details are provided in the caption) were performed under the null hypothesis of Normally distributed data, and all the results we obtained prevent from rejecting such hypothesis.

\section{Empirical data: the inverse problem}
		
The goal of this section is to show that some of the features displayed in correlation matrix spectra of factor models are actually present in empirical spectra of financial correlation matrices too. In particular, our goal is to show that the empirically observed eigenvalue bulks, as the ones shown in Figure \ref{SP_FTSE_spectra}, cannot be regarded as a consequence of pure noise at all. Indeed, by suitably filtering empirical data, it is possible to show a peak separation similar to those shown in Figure \ref{Factor_models_spectra1} and Figure \ref{Factor_models_spectra2}. \\
Starting from the two data sets already introduced in a previous section ($396$ assets from the S$\&$P500 Index and $243$ assets from the FTSE350 Index), we shall restrict our attention only to a relatively small number of properly chosen assets. This will be done in order to ideally recreate, as best as possible, the conditions under which the previously discussed eigenvalue bulks emerge from factor models. We shall attempt to empirically recreate the block-diagonal correlation matrix in \eqref{covblock2}, which, when a single cluster is considered, yields the eigenvalue spectrum of equation \eqref{thspectrum3}. Let us rewrite that matrix in this specific case:

\begin{eqnarray} \label{covblock3}
\mathbf{C} = \left ( \begin{array}{cc}
		\tilde{\mathbf{E}}^{\bar{(N)}} & 0 \\
		0 & \mathbf{I}_{N-\bar{N}} \\
	\end{array} \right),
\end{eqnarray}
recalling that $\tilde{E}^{\bar{(N)}}_{ij} = \rho$ for $i \neq j$ and $\tilde{E}^{\bar{(N)}}_{ii} = 1$. \\
In order to reproduce the structure in \eqref{covblock3}, we start from the empirical covariance matrices (let us denote them as $\mathbf{c}$, according to the previously adopted notation) of our data sets and apply the following procedure.

\begin{itemize}
\item We first identify a small cluster of $\bar{N}$ strongly mutually correlated assets. If we denote the corresponding set of indices as $I_U$, then we have
\begin{equation} \label{cluster}
c_{ij} \geq \rho_U \ ,\ \ \ \ i,j \in I_U
\end{equation}
for some threshold value $\rho_U > 0$.
\item Then, all assets which are weakly correlated to the elements in the cluster are pointed out. Amongst those, only the ones with small mutual correlations are retained. By grouping their indices in another set $I_D$, we can write
\begin{eqnarray} \label{uncorr}
| c_{ij} | &\leq& \rho_D^{\prime} \ ,\ \ \ \ i \in I_U, \ j \in I_D \\ \nonumber
| c_{kl} | &\leq& \rho_D^{\prime \prime} \ , \ \ \ k,l \in I_D, \ k \neq l
\end{eqnarray}
for some threshold values $\rho_D^{\prime}, \rho_D^{\prime \prime} \in (0, \rho_U)$ such that $\rho_D^{\prime} \leq \rho_D^{\prime \prime}$. \\
The first condition in \eqref{uncorr} is meant to reproduce the zero off-diagonal blocks in \eqref{covblock3}, while the second one is meant to reproduce the identity matrix in the right-lower block.
\item If we now redefine $N$ to be the total number of stocks in $I_U$ and $I_D$, so that $I_D$ contains $N-\bar{N}$ elements, and we properly sort them, then the approximation to \eqref{covblock3} is given as follows
\begin{equation} \label{covblockest}
\mathbf{c} = \left ( \begin{array}{cc}
		\mathbf{c}_{I_U} & \mathbf{c}_{I_U,I_D} \\
		\mathbf{c}_{I_D,I_U} & \mathbf{c}_{I_D} \\
	\end{array} \right),
\end{equation}
where $\mathbf{c}_{I_U}$ and $\mathbf{c}_{I_D}$ are square matrices (of dimensions $\bar{N}$ and $N-\bar{N}$ respectively) containing the correlation matrix elements pertaining to the two sets $I_U$ and $I_D$. On the other hand, the $\mathbf{c}_{I_U,I_D}$ matrix ($\mathbf{c}_{I_D,I_U}$ being its transpose) contains the ``interaction'' terms between the two sets.
\end{itemize}
The goal of such a construction is to empirically make contact with the spectrum in \eqref{thspectrum3}. As a matter of fact, for suitably chosen threshold values $\rho_U$, $\rho_D^{\prime}$ and $\rho_D^{\prime \prime}$, we expect the eigenvalue spectrum of the $\mathbf{c}$ matrix in \eqref{covblockest} to be the noise-dressed version of the one in \eqref{thspectrum3}. In particular, small values of $\rho_D^{\prime}$ and $\rho_D^{\prime \prime}$ should guarantee the $\mathbf{c}_{I_D}$ block to yield $N-\bar{N}$ eigenvalues close to one. On the other hand, the block $\tilde{\mathbf{E}}^{(\bar{N})}$ in \eqref{covblock3} yields $\bar{N}-1$ small eigenvalues equal to $1-\rho$ and a large one equal to $\bar{N} \rho + (1-\rho)$. Now, it is reasonable to assume $\rho$ to be equal to the average mutual correlation between the assets in $I_U$

\begin{equation} \label{meanrho}
\rho = \frac{1}{\bar{N}(\bar{N}-1)} \sum_{i \neq j} [c_{I_{U}}]_{ij}
\end{equation}
and to suppose that $\mathbf{c}_{I_U}$ will produce $\bar{N}-1$ eigenvalues close to this value.
\begin{figure}
\begin{center}
\includegraphics[width=2.9 in, height=2.0 in]{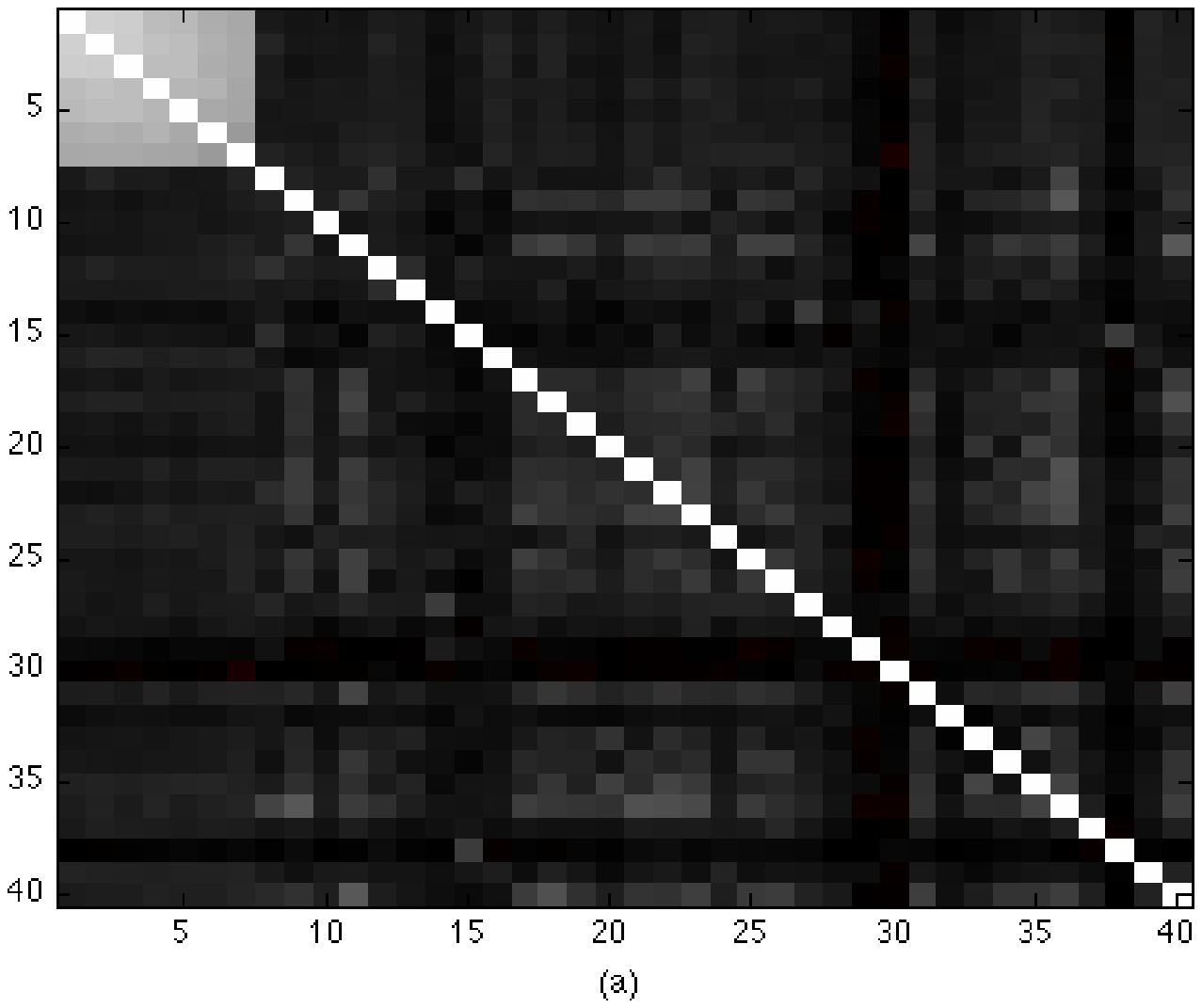} 
\hspace{0.1 in} 
\includegraphics[width=2.9 in, height= 2.0 in]{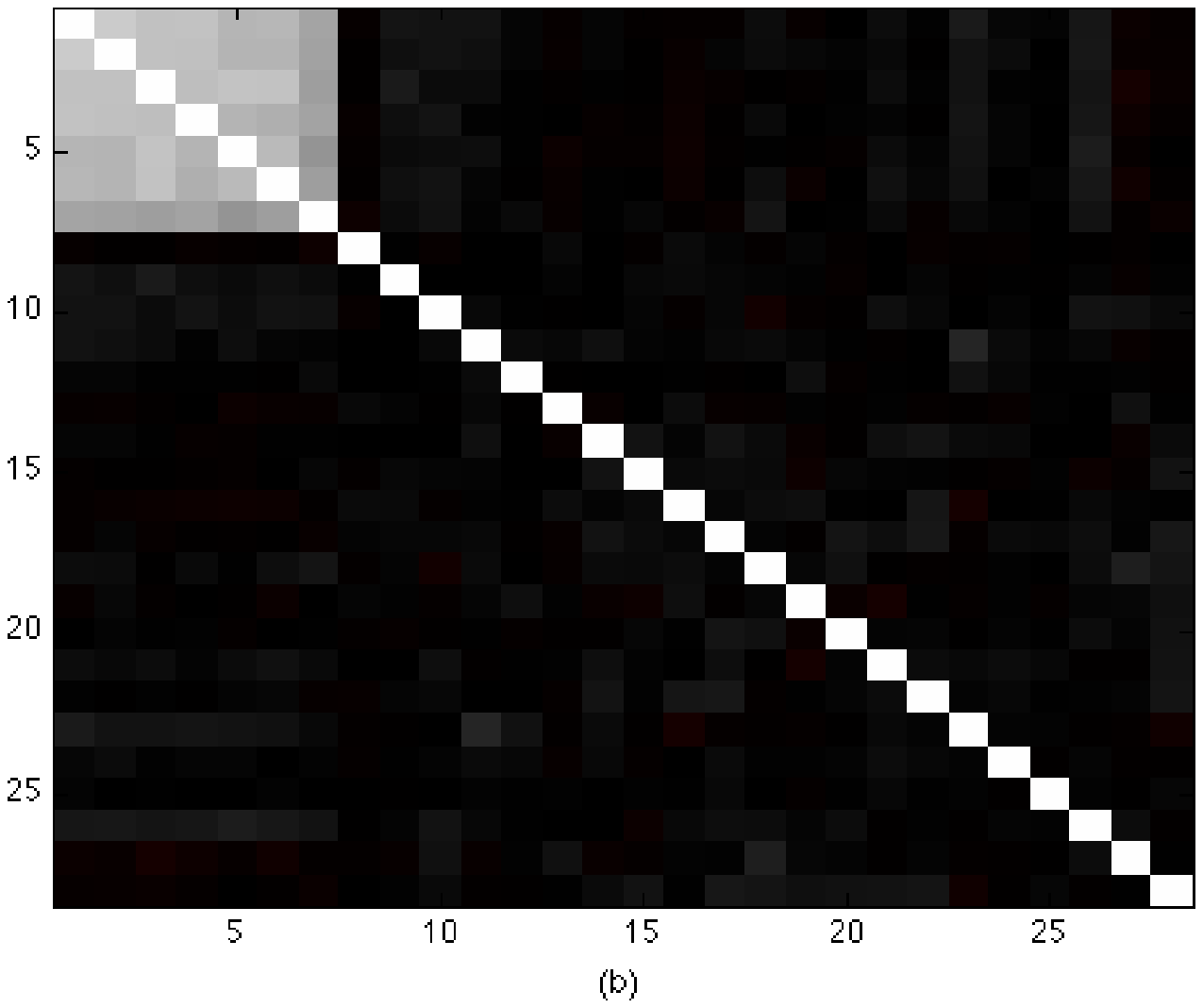} 
\vspace{0.01 in}
\vspace{0.1 in}
\includegraphics[width=2.9 in, height= 2.0 in]{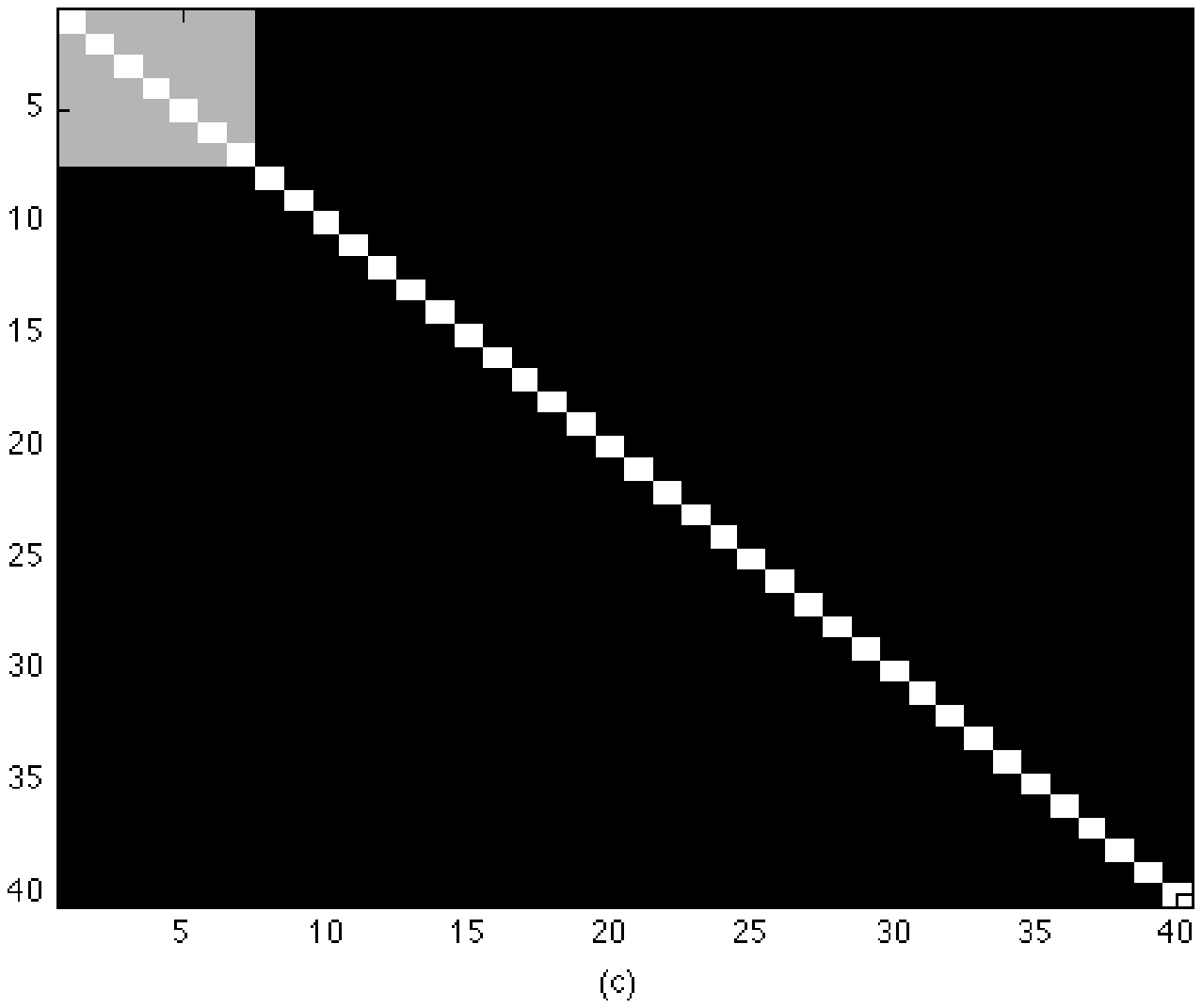}
\hspace{0.1 in} 
\includegraphics[width=2.9 in, height= 2.0 in]{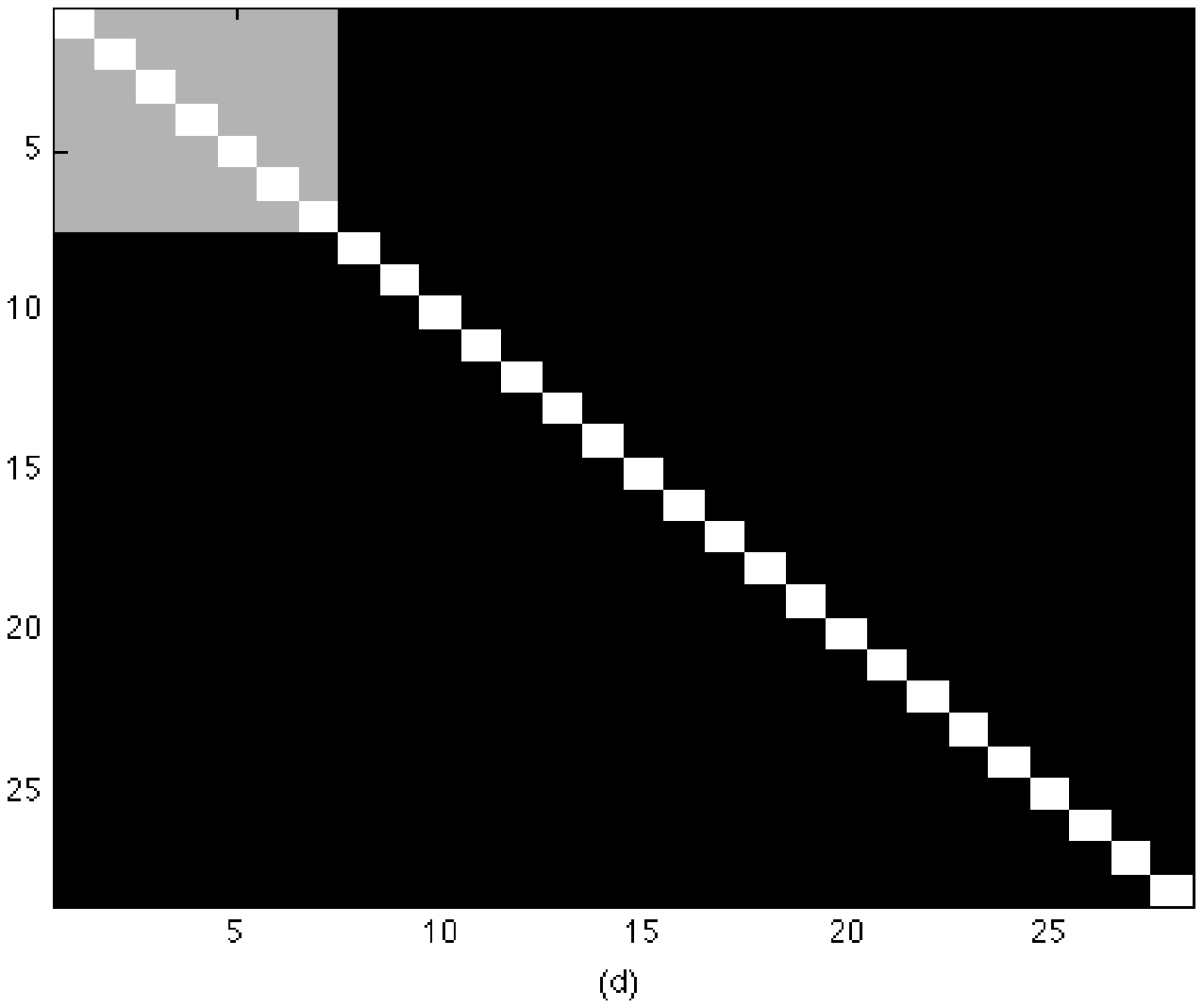} 
\caption{Graphical representation of correlation matrices. (a) $40 \times 40$ correlation matrix for the selected returns belonging to the S$\&$P500 Index. (b) $28 \times 28$ correlation matrix for the selected returns belonging to the FTSE350 Index. In (a) and (b) the stocks have been sorted in order to highlight the cluster structure. (c)-(d) Model correlation matrices corresponding to the cases shown in (a) and (b), respectively. In all plots, white diagonal blocks correspond to ones, while black ones correspond to zeroes. Gray shadings are intermediate values. As already explained in the main text, the gray shadings in (c) and (d) correspond to $\rho = 0.712$ and $\rho = 0.707$, respectively. The presence of unresolved structures in (a) and, less evident, in (b), suggests that the model matrix in \eqref{covblock3}, depicted in (c) and (d), does not fully capture all empirical features. } \label{corrcolormap}
\end{center}
\end{figure}
\\ In the following we present and discuss the results we obtained applying this procedure to our datasets. In the case of the S$\&$P500 Index, we identified a cluster made of $\bar{N}_{\mathrm{S}\&\mathrm{P}} = 7$ strongly mutually correlated assets ($\rho_{\mathrm{S}\&\mathrm{P}} = 0.712$, with $\rho_{\mathrm{S}\&\mathrm{P}}$ computed as in \eqref{meanrho}), all of which happen to belong to the energy sector. We then identified a group of $33$ stocks, belonging to various sectors, which satisfy the previously described requirements: a small mutual correlation (mean value $= 0.099$) and a small correlation with the $\bar{N}$ elements in the cluster (mean value $= 0.096$). So, all in all we have $N_{\mathrm{S}\&\mathrm{P}} = 40$. Analogously, also in the FTSE350 Index case we were able to identify a cluster made of $\bar{N}_{\mathrm{FTSE}} = 7$ highly mutually correlated stocks ($\rho_{\mathrm{FTSE}} = 0.707$), all corresponding to investment trusts. In this case, however, we only found $21$ more stocks (so that $N_{\mathrm{FTSE}} = 28$) satisfying the aforementioned requirements (mean value of mutual correlation $= 0.015$, mean value of correlations with elements in the cluster $= 0.014$). In Figure \ref{corrcolormap}, graphical representations of the empirical correlation matrices we obtained, and a comparison to the theoretically expected ones, are shown. As can be seen by direct inspection, the correlation matrix we obtain in the FTSE350 Index case is remarkably similar to the one in \eqref{covblock3}, whereas  the one we obtain for the S$\&$P500 Index has some further inner structure as a consequence of the much higher mean correlations.
\begin{figure}
\begin{center}
\includegraphics[width=2.9 in, height=2.0 in]{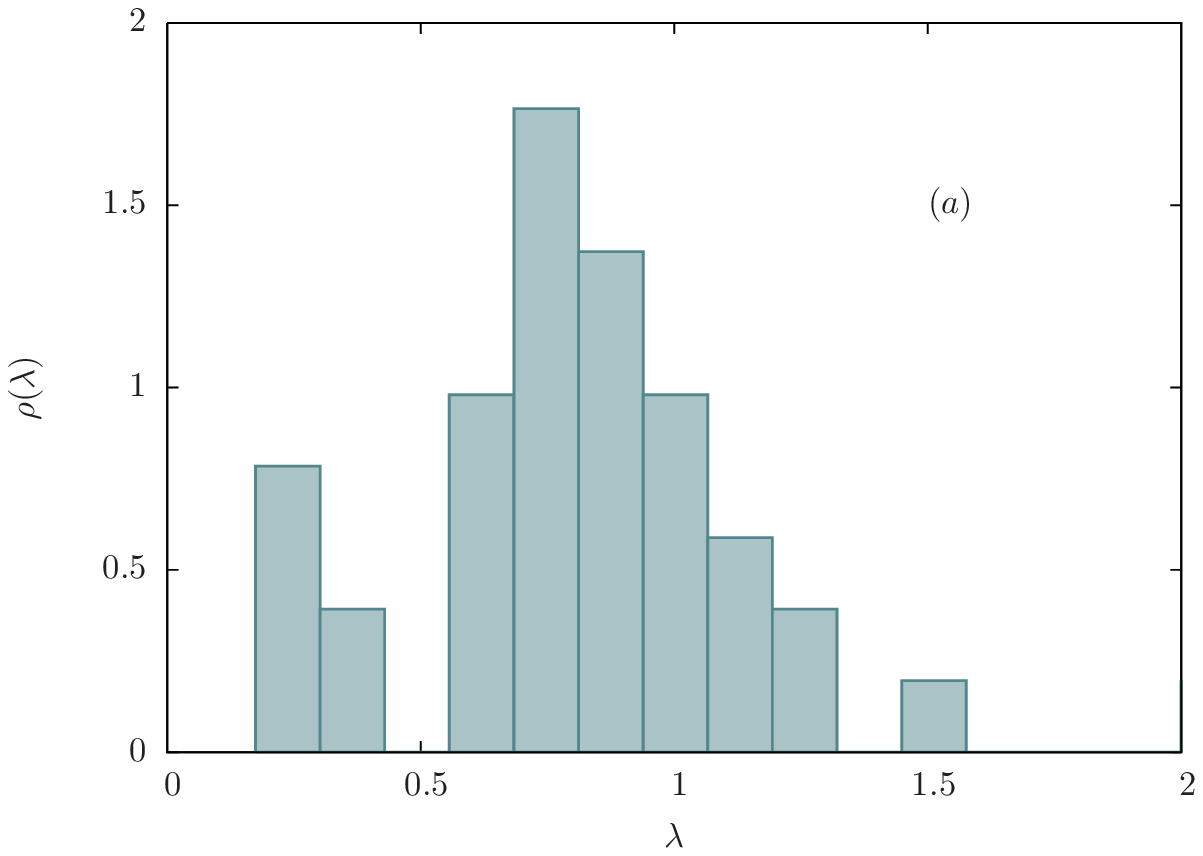} 
\hspace{0.1 in} 
\includegraphics[width=2.9 in, height= 2.0 in]{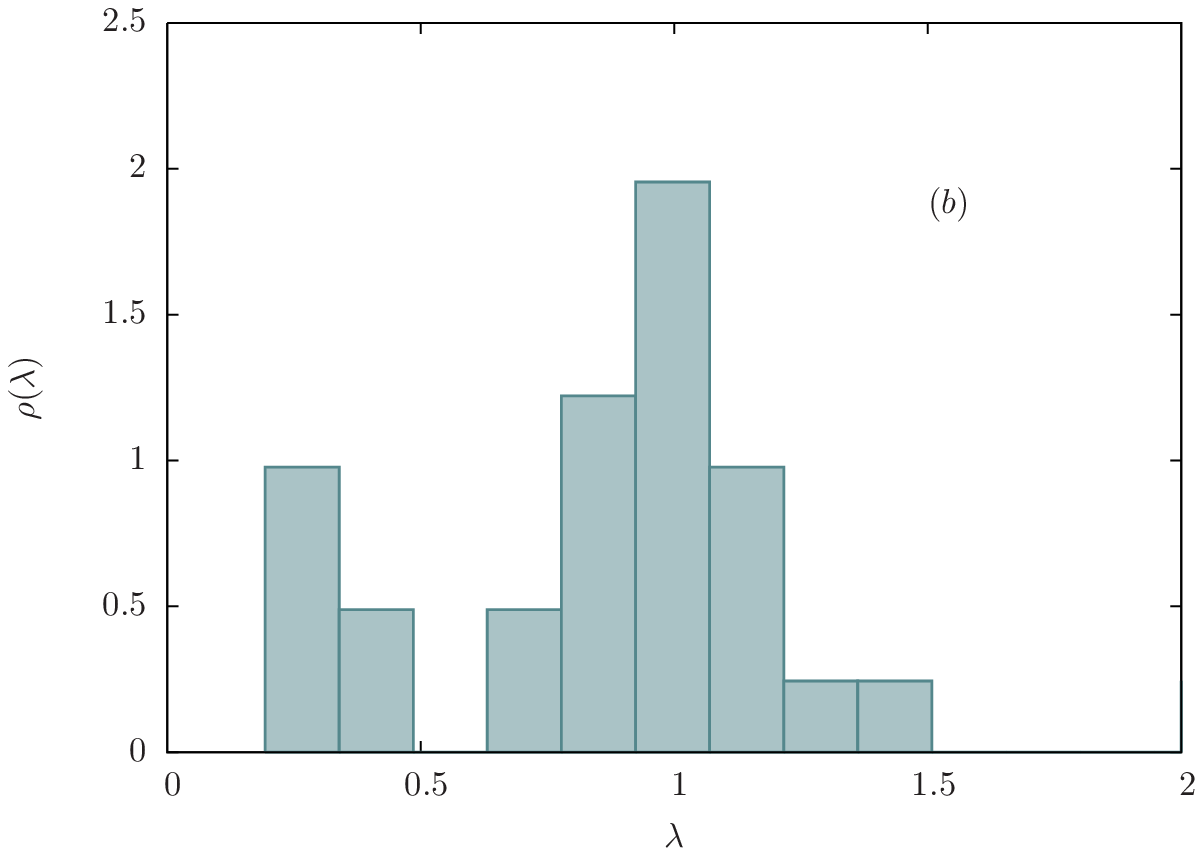} 
\vspace{0.1 in}
\includegraphics[width=2.9 in, height= 2.0 in]{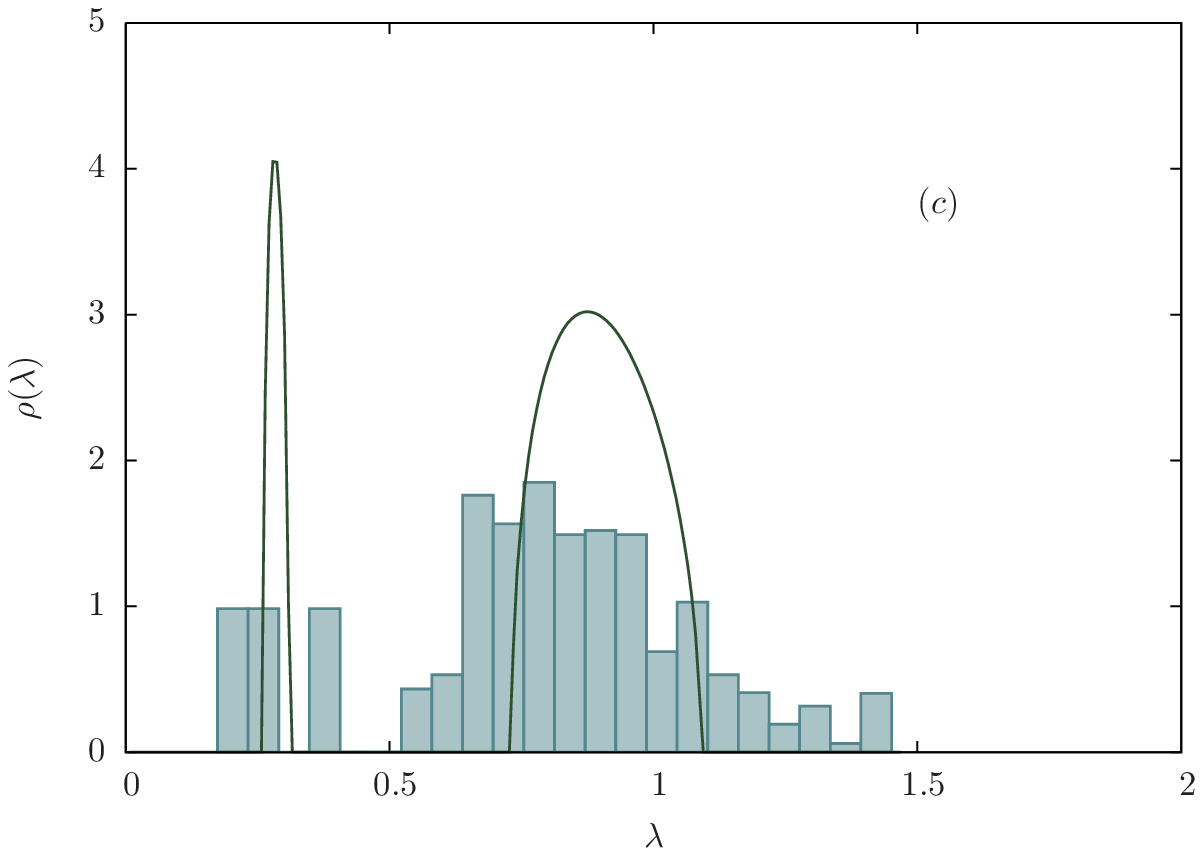}
\hspace{0.1 in} 
\includegraphics[width=2.9 in, height= 2.0 in]{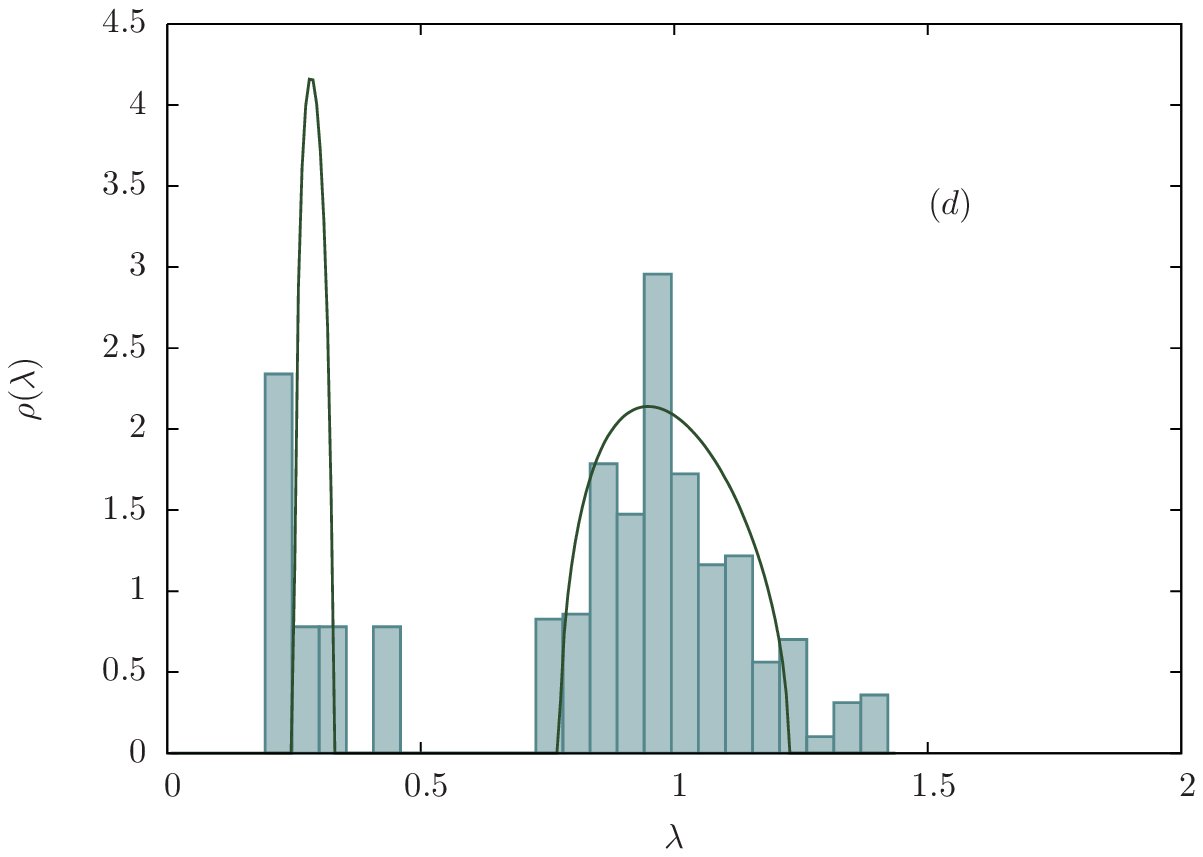} 
\vspace{0.1 in}
\includegraphics[width=2.9 in, height= 2.0 in]{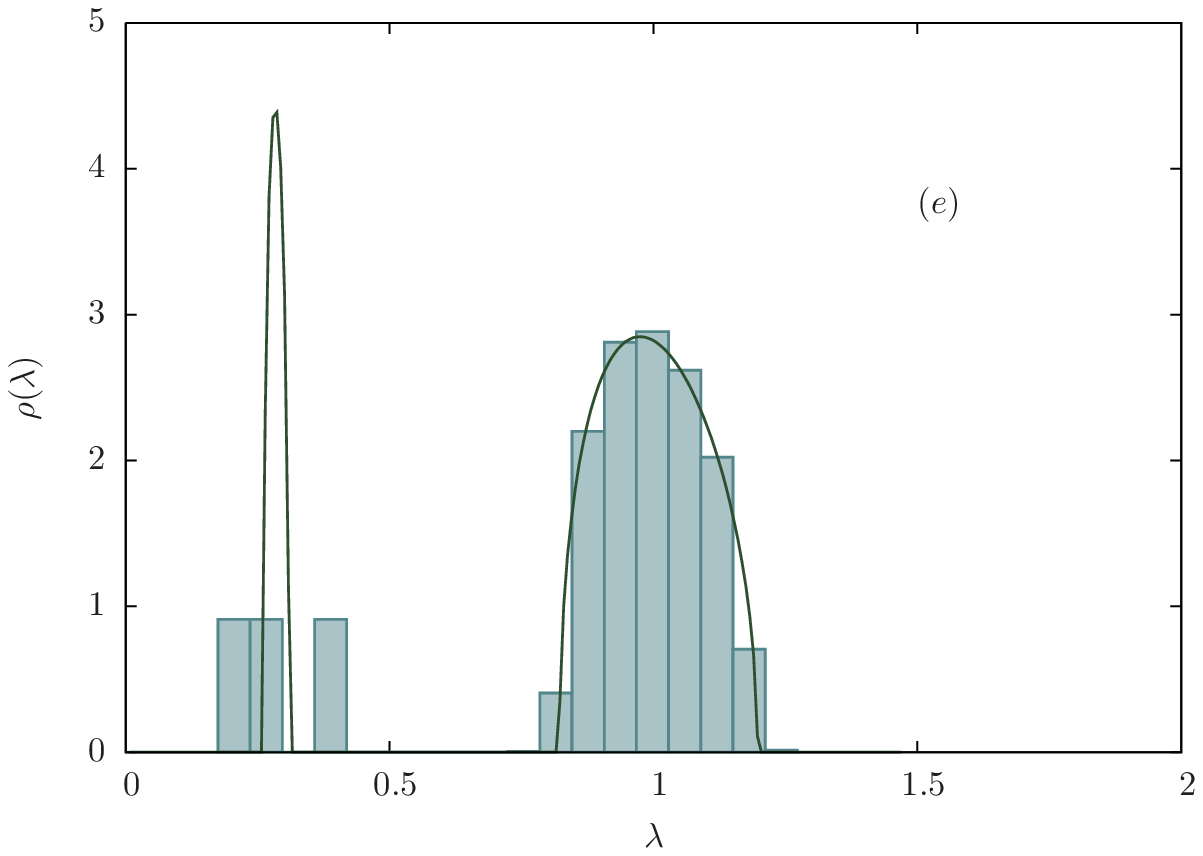}
\hspace{0.1 in} 
\includegraphics[width=2.9 in, height= 2.0 in]{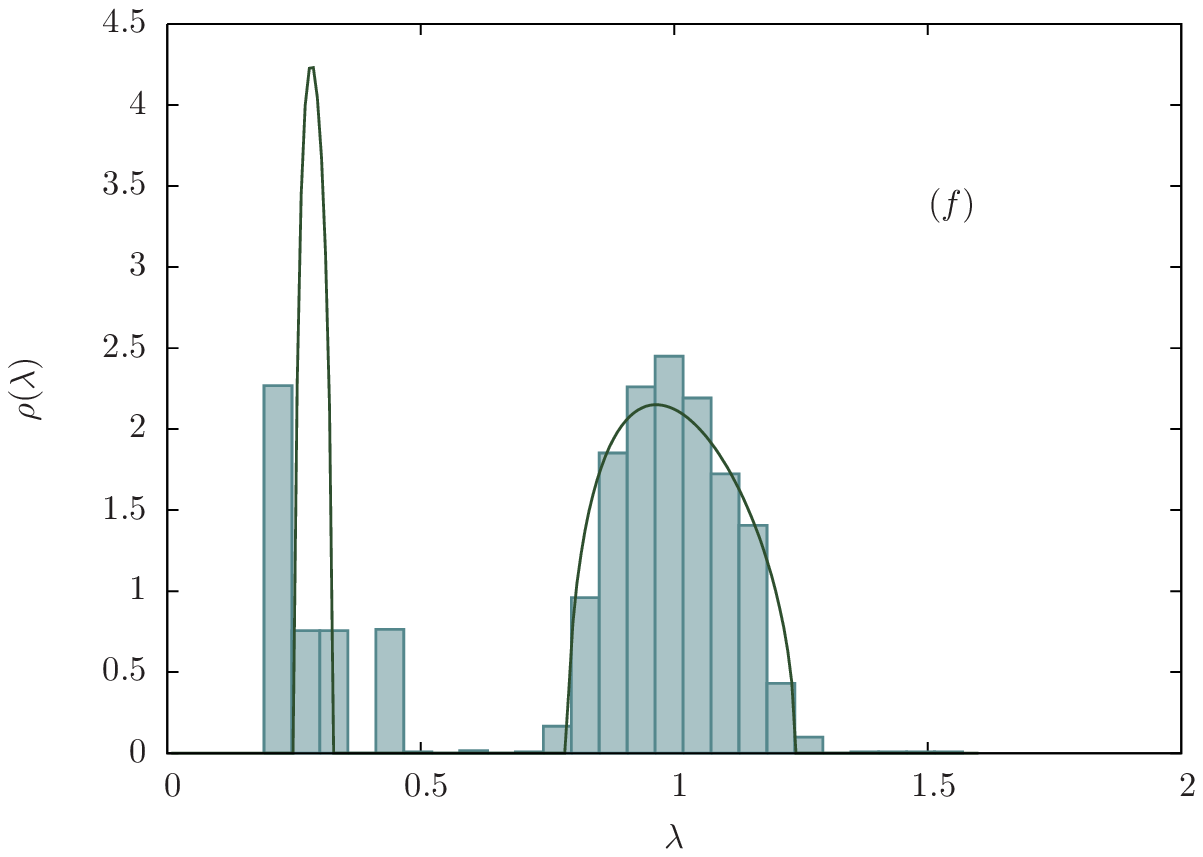} 
\caption{(a)-(b) Eigenvalue densities of the correlation matrices represented in Figure \ref{corrcolormap} (a)-(b), relative to the S$\&$P500 and FTSE350 Indices, respectively. In both cases, one can clearly distinguish two well separated bulks, while the largest eigenvalues have not been plotted for better visualization (see main text for further explanation). (c)-(d) Comparison between the theoretically expected spectra derived via equation \eqref{meq} and the empirical ones. The latter have been modified with respect to (a) and (b) according to the following approach. A bootstrap random sampling (100 iterations) has been performed on the weakly correlated subsets of stocks, picking 30 stocks out of 33 in the S$\&$P500 Index case and 18 out of 21 in the FTSE350 Index case. The presence of undetected structures (see Figure \ref{corrcolormap} (a)-(b) and main text) leads to a poor agreement between data and theory. The values and weights of the eigenvalues used to plot the theoretical density obtained from equation \eqref{rhofromM} are detailed in \eqref{lambdaval1}. (e)-(f) As in (c)-(d) but with weakly correlated data reshuffled before bootstrap, leading to a much better agreement between data and theory. The reference eigenvalue for the bulks on the right is now assumed to be equal to one (see main text).} \label{Empirical_spectra}
\end{center}
\end{figure}
\\ In Figure \ref{Empirical_spectra} the eigenvalue spectra we obtained from the previously discussed correlation matrices (the ones reported in Figure \ref{corrcolormap} (a)-(b)) are shown. In particular, in Figure \ref{Empirical_spectra} (a)-(b) we plot the spectra obtained, respectively, from the S$\&$P500 and FTSE350 correlation matrices constructed according to the clustering procedure outlined previously. In both cases, two distinct eigenvalue bulks can be noticed. The smaller bulks on the left are made of 6 eigenvalues, and, since we have $\bar{N}_{\mathrm{S}\&\mathrm{P}} = \bar{N}_{\mathrm{FTSE}} = 7$, this is in agreement with the prediction given by equation \eqref{thspectrum3} for $\bar{N} = 7$. Also, in the FTSE350 Index case (Figure \ref{corrcolormap} (b)) the larger eigenvalue bulk around one is made of $N_{\mathrm{FTSE}}-\bar{N}_{\mathrm{FTSE}} = 21$ eigenvalues, which is again in agreement with \eqref{thspectrum3}, while the largest eigenvalue in the spectrum (not shown in the plot) is equal to 5.235, remarkably close to the prediction given by $\bar{N}_{\mathrm{FTSE}} \ \rho_\mathrm{FTSE} + (1-\rho_{\mathrm{FTSE}}) = 5.242$ (see again equation \eqref{thspectrum3}). On the other hand, the spectrum relative to the S$\&$P500 Index yields two large eigenvalues (not shown in Figure \ref{Empirical_spectra} (b)) equal to 3.552 and 6.483, and neither value is in agreement to the large eigenvalue prediction $N_{\mathrm{S}\&\mathrm{P}} \ \rho_{\mathrm{S}\&\mathrm{P}} + (1 - \rho_{\mathrm{S}\&\mathrm{P}}) = 5.272$. Such discrepancy is due to the unresolved correlation structure in the empirical S$\&$P matrix (see Figure \ref{corrcolormap} (a)), which gives rise to additional sub-clusters. \\
In Figure \ref{Empirical_spectra} (c)-(d) the two eigenvalue bulks we just discussed are fitted with the eigenvalue density deriving from the second equation in \eqref{rhofromM} when applied to the solution of \eqref{meq}, \emph{i.e.} the moment generating function $m_{\mathbf{c}}$ of the noise-dressed version of a correlation matrix $\mathbf{C}$ with degenerate eigenvalues. In both cases we consider correlation matrices with two degenerate eigenvalues in order to try to fit the two main bulks. The smaller eigenvalue $\Lambda_1$, responsible for the emergence of the smaller bulks on the left, is assumed to be equal to $1 - \rho$, accordingly to equation \eqref{thspectrum3}. So, in the two different cases we analyzed we have

\begin{eqnarray} \label{lambdaval1}
\Lambda_{1_{\mathrm{S}\&\mathrm{P}}} = 1 - \rho_{\mathrm{S}\&\mathrm{P}} = 0.288 &,& \ \ \ \ w_{1_{\mathrm{S}\&\mathrm{P}}} = \frac{\bar{N}_{\mathrm{S}\&\mathrm{P}} - 1}{N_{\mathrm{S}\&\mathrm{P}} - 2} \\ \nonumber
\Lambda_{1_{\mathrm{FTSE}}} = 1 - \rho_{\mathrm{FTSE}} = 0.293 &,& \ \ \ \ w_{1_{\mathrm{FTSE}}} = \frac{\bar{N}_{\mathrm{FTSE}} - 1}{N_{\mathrm{FTSE}} - 1}
\end{eqnarray}
where the two slightly different weights are justified by the previously mentioned fact that in the S$\&$P case there are two isolated eigenvalues which separate from the main bulks, while in the FTSE case there is only one such eigenvalue. On the other hand, the larger eigenvalue $\Lambda_2$, which according to \eqref{thspectrum3} should be exactly equal to one, is assumed to be equal to the empirical mean value of the main bulks on the right in Figure \ref{Empirical_spectra} (c)-(d). These are found to be

\begin{equation} \label{lambdaval2}
\Lambda_{2_{\mathrm{S}\&\mathrm{P}}} = 0.887 \ , \ \ \ \ \Lambda_{2_{\mathrm{FTSE}}} = 0.997
\end{equation}
and one might notice that, again, the value obtained in the FTSE case is in excellent agreement with the theoretically expected one. So, all in all, the curves drawn in Figure \ref{Empirical_spectra} (c)-(d) are obtained from the values in equations \eqref{lambdaval1} and \eqref{lambdaval2}. Such curves, as already mentioned, are fitted to the empirical spectra. However, a bootstrap approach was adopted in order to improve the statistics. More specifically, for each bootstrap iteration a random sampling on the weakly correlated stocks was performed, picking 30 out of 33 in the S$\&$P case and 18 out of 21 in the FTSE case. On the contrary, the stocks forming the highly correlated clusters were always kept (thus keeping the eigenvalue bulks on the left almost unchanged). As it can be seen in Figure \ref{Empirical_spectra} (c)-(d) the agreement between theory and prediction is very poor. This is essentially due to the additional correlation structures in the empirical correlation matrices (see Figure \ref{corrcolormap}), which are neglected in the model matrix \eqref{covblock3} and in its eigenvalue spectrum \eqref{thspectrum3}. All the bulks displayed in Figure \ref{Empirical_spectra} (c)-(d) appear to be ``smeared'' versions of their theoretical counterparts, even the small ones relative to the eigenvalues in \eqref{lambdaval1}. Interestingly, this shows that inhomogeneities in correlation structures have quite an impact on eigenvalue spectra even on a ``small scale'' (let us recall that $\bar{N}_{\mathrm{S}\&\mathrm{P}} = \bar{N}_{\mathrm{FTSE}} = 7$). \\
In Figure \ref{Empirical_spectra} (e)-(f) the same fit as the one just discussed is performed, the only difference being that an additional random reshuffling of the returns is performed on the bootstrapped assets. Such an operation is meant to destroy all possible correlations, and this leads to a quite good agreement between data and predictions on the bulks on the right (the theoretical densities being now computed with $\Lambda_2 = 1$, accordingly to equation \eqref{thspectrum3}). This essentially confirms that the substantial deviations shown in Figure \ref{Empirical_spectra} (c)-(d) can entirely be imputed to the unresolved cluster structures in the empirical correlation matrices. The same kind of analyses (bootstrap and reshuffling) were not performed on the stocks belonging to the correlated clusters because of their very small number. \\
All in all, the previous observations definitely suggest that the empirically observed eigenvalue bulks cannot be regarded as a consequence of the noisiness in financial correlation matrices. On the contrary, in the light of the previous discussions it could be conjectured that bulks emerge from the interplay of several cluster structures like the ones we isolated (see Figure \ref{corrcolormap}) \cite{Burda2004_1, Burda2004_2}.

\section{Summary and conclusions}

Let us now summarize the main messages in the paper.
\begin{itemize}
\item Several rough but useful results about spectral properties of financial correlation matrices, such as the position of large non-degenerate eigenvalues, can be inferred by a clever application of the direct problem (see Section II A). This only involves algebraic calculations, namely the solution of suitable secular equations. This approach can be used either when the cluster structure is known \emph{a priori}, or when there are good reasons to assume a certain correlation structure. Combining the direct analysis with Monte Carlo simulations can provide a clear picture in a number of situations, avoiding the analytical difficulties of Random Matrix Theory, and keeping the finite-sized nature of the problem. Typically, one wishes to reproduce observed spectra starting from a factor model, and this can be done as follows.
\begin{enumerate}
\item Identify the cluster structure in the dataset under analysis, using clustering algorithms \cite{Aldenderfer}.
\item Estimate the average correlations within clusters.
\item Build a theoretical, ``mean field'', matrix model $\mathbf{C}$ from the above estimates.
\item Run Monte Carlo simulations of the matrix model. 
\item Compare the outcome of the simulation to the empirical spectrum.
\end{enumerate}
If the comparison is statistically satisfactory, the matrix model $\mathbf{C}$ can be retained and used for further analyses, such as portfolio selection. If not, the model is to be refined by abandoning the mean field assumption, at least for some cluster interactions.

\item As far as the largest eigenvalue is concerned, its distribution is not Tracy-Widom, but Normal \cite{Tracy, Paul}. Moreover, this distribution cannot be derived from the thermodynamic limit formula \eqref{meq}. In fact, such an eigenvalue is typically non-degenerate and its weight in a diagrammatic expansion of the Green's function would vanish as $1/N$, for $N \rightarrow \infty$.

\item For factor models, the bulks in empirical eigenvalue spectra come as the noise-dressed version of degenerate eigenvalues. Thus, such bulks encode the information on the cluster structure of the empirical correlation matrix $\mathbf{c}$, and this can be evidenced by means of proper clustering methods, as done in Section III.
\end{itemize}
While there would be no difficulty in studying non-Gaussian multivariate models by means of Monte Carlo simulations, the analytical results presented in Section II B and in the Appendix cannot be easily generalized. In fact, the integrals needed to calculate $\mathbf{g}_{\mathbf{c}}$ in \eqref{Green} in the Gaussian case can be exactly obtained by virtue of Wick's theorem, whereas different stochastic models would require painful calculations. \\ 
The diagrammatic method outlined in the Appendix allows, in principle, for the exact evaluation of the Green's function $\mathbf{g}_{\mathbf{c}}$ for any finite size $N \times T$, as a function of $N$ and $T$. Nevertheless, this is a series of $1/z$ powers, whose convergence properties would be interesting to investigate in the near future.

\begin{acknowledgements}
Simone Alfarano acknowledges financial support by the Spanish Ministry of Science and Innovation from research project ECO2008-00510, and by Universitat Jaume I - Bancaixa from research project P11A2009-09.
Enrico Scalas is grateful to Universitat Jaume I for the financial support received from their Research Promotion Plan 2010 during his scientific visit in Castell\'on, where the authors met to work on this paper. Giacomo Livan wishes to thank G. Montagna and O. Nicrosini for their kind support and helpful suggestions. The authors are indebted to Z. Burda for useful discussions.
\end{acknowledgements}

\appendix
\section{Diagrammatic method}
\begin{figure}
\begin{center}
\includegraphics[width=3.9 in, height=1.9 in]{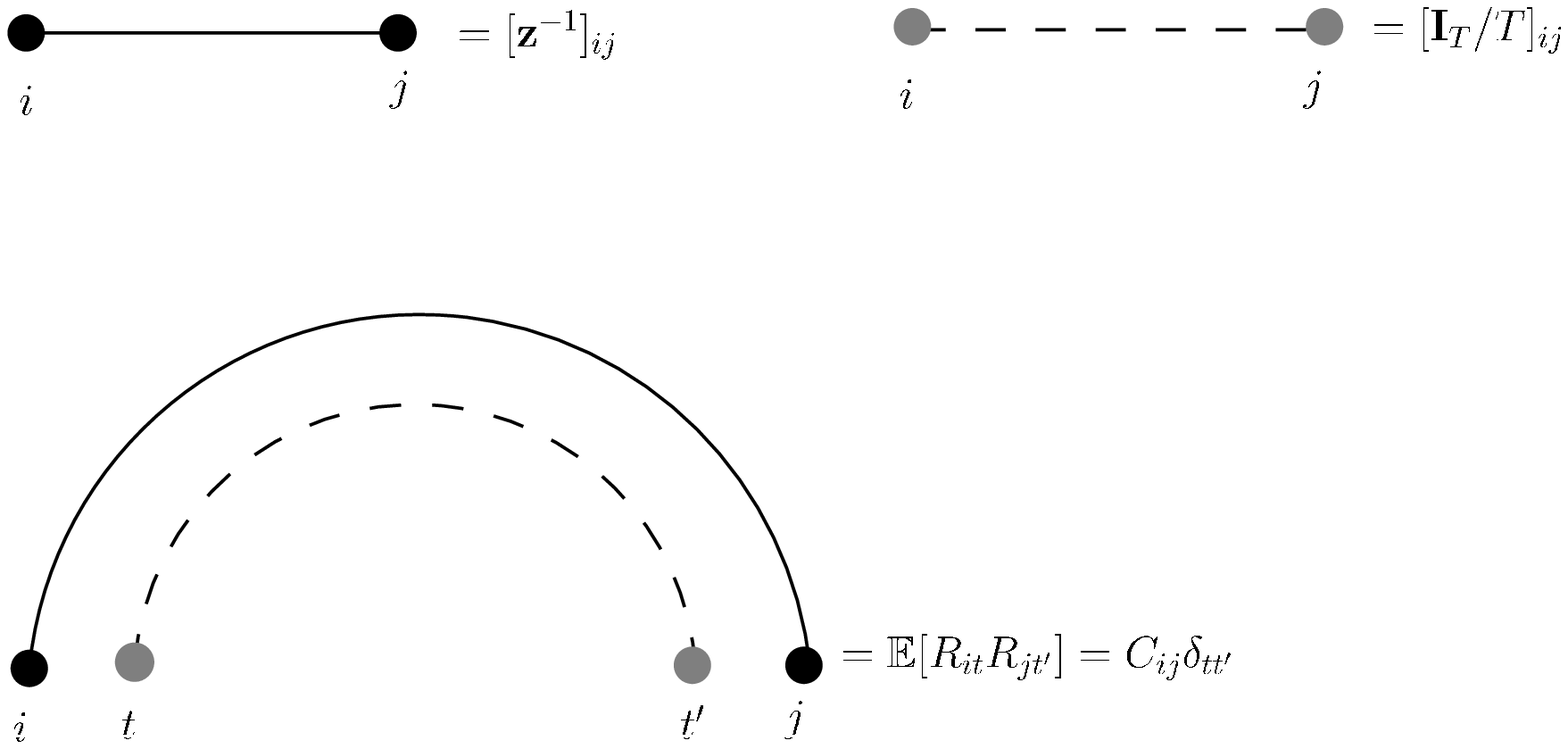} 
\caption{Representations of the $\mathbf{z}^{-1}$ matrix, the $\mathbf{I}_T/T$ matrix and the propagator.} \label{Symbols}
\end{center}
\end{figure}

\begin{figure}
\begin{center}
\includegraphics[width=4.4 in, height= 5.4 in]{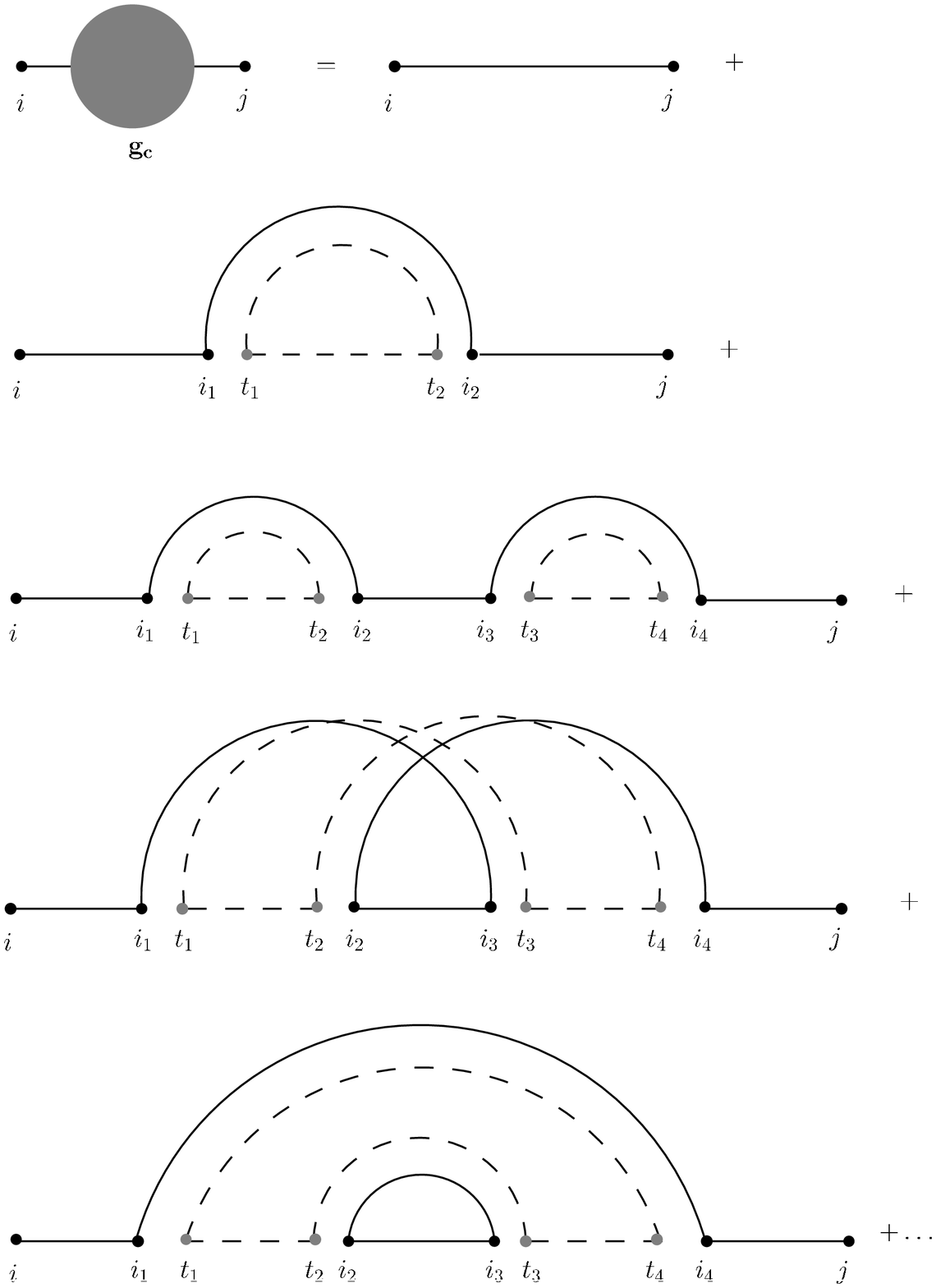} 
\caption{Diagrammatic expansion of the Green's function (represented as a circle within two $N$-type matrix indices). The diagrams on the first two lines represent the contributions of the $\mathbf{z}^{-1}$ matrix and the two-point correlation function (see Figure \ref{Symbols}) respectively. The other diagrams on the following lines represent the four-point correlation function, and in particular they represent the different contributions which arise by applying Wick's Theorem as in equation \eqref{4point}. The diagram on the fourth line is non-planar, and its contribution to the expansion is negligible when the thermodynamic limit \eqref{thermlim} is taken (see the main text for further clarifications). Higher order diagrams are not represented.} \label{GreenDiag}
\end{center}
\end{figure}

\begin{figure}
\begin{center}
\includegraphics[width=4.0 in, height=3.0 in]{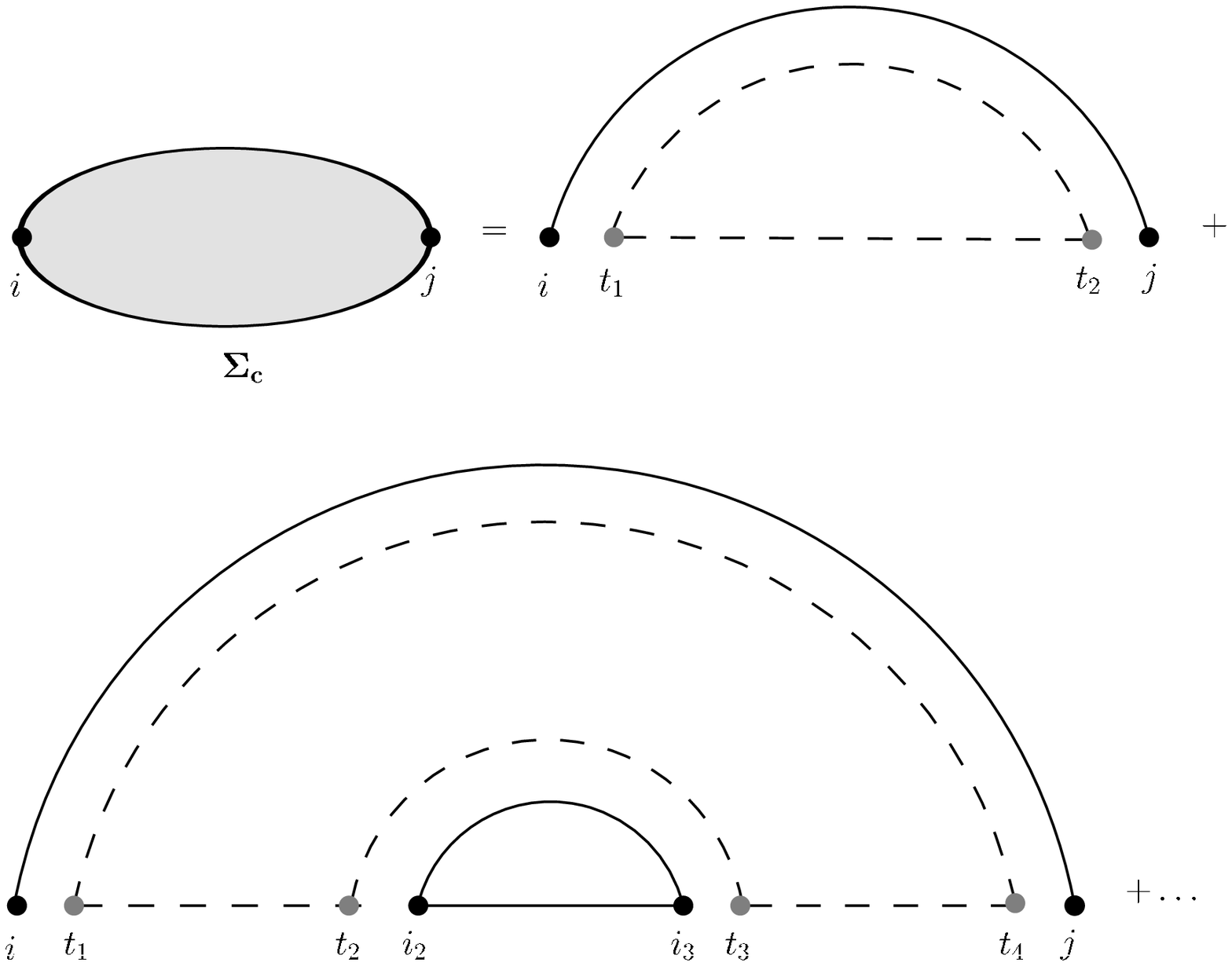} 
\caption{Representation of the self energy as the generating function of one-line-irreducible (rainbow) diagrams.} \label{SelfEnergy}
\end{center}
\end{figure}

\begin{figure}
\begin{center}
\includegraphics[width=4.0 in, height=2.9 in]{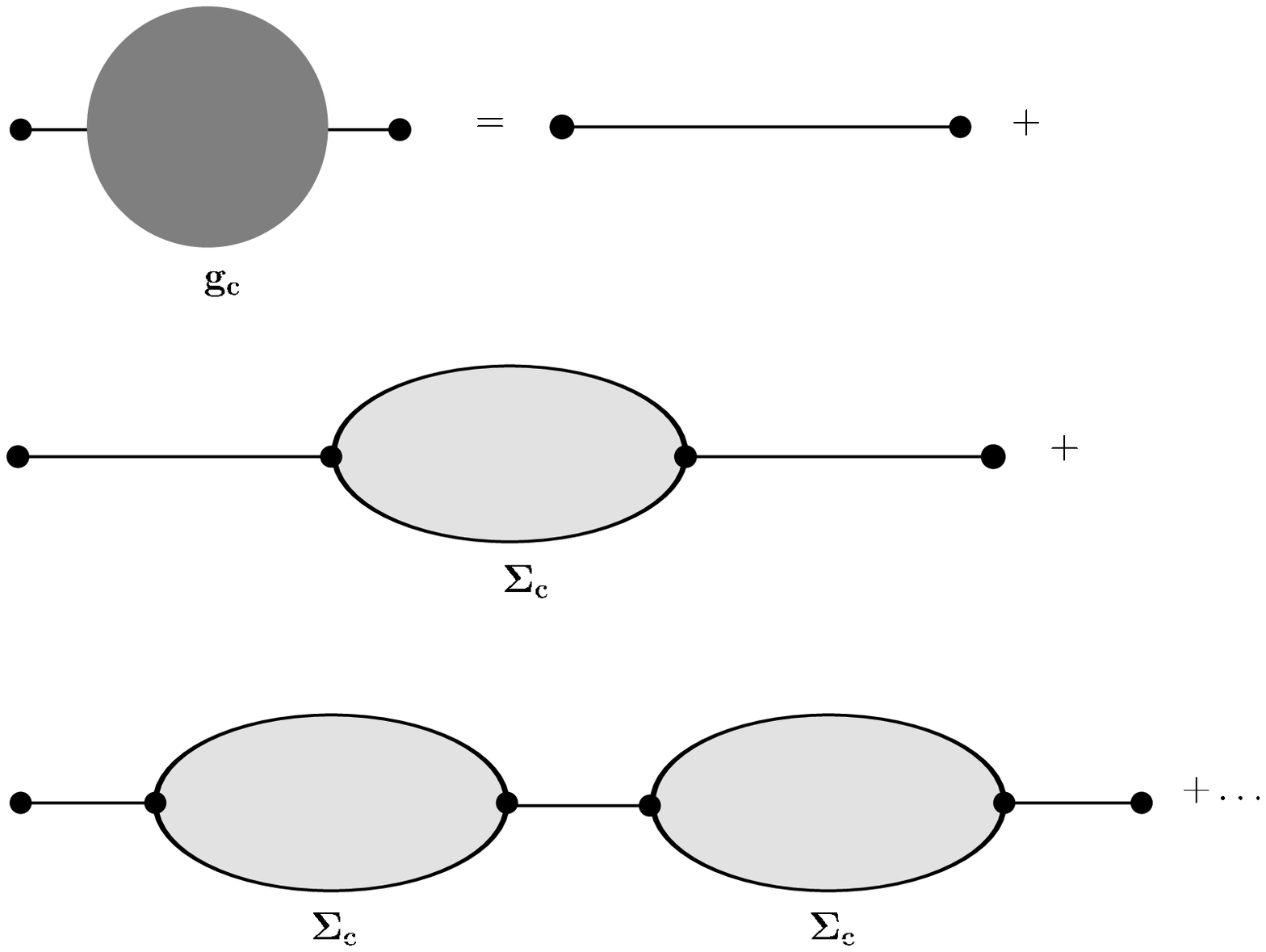} 
\caption{Expansion of the Green's function in terms of the self-energy.} \label{GreenS}
\end{center}
\end{figure}

For the sake of completeness, in this Appendix we replicate the derivation, already detailed in \cite{Burda2004_1}, of equations \eqref{mMrel} and \eqref{confmap}. The starting point is to expand the second Green's function in \eqref{Green} (let us now pose $\mathbf{z} = z \mathbf{I}_N$, with $z \in \mathbb{C}$)

\begin{eqnarray} \label{Greenexp}
\mathbf{g}_{\mathbf{c}}(z) &=& \mathbb{E} \left [ (\mathbf{z} - \mathbf{c})^{-1} \right ] = \mathbf{z}^{-1} +
\mathbb{E} \left [ \mathbf{z}^{-1} \mathbf{c} \ \mathbf{z}^{-1} \right ] + \mathbb{E} \left [ \mathbf{z}^{-1} \mathbf{c} \ \mathbf{z}^{-1} \mathbf{c} \ \mathbf{z}^{-1} \right ] + \ldots \\ \nonumber
&=& \mathbf{z}^{-1} + \mathbb{E} \left [ \mathbf{z}^{-1} \mathbf{R} \ \frac{\mathbf{I}_T}{T} \ \mathbf{R}^{\mathrm{T}} \mathbf{z}^{-1} \right ] + \mathbb{E} \left [ \mathbf{z}^{-1} \mathbf{R} \ \frac{\mathbf{I}_T}{T} \ \mathbf{R}^{\mathrm{T}} \mathbf{z}^{-1} \mathbf{R} \ \frac{\mathbf{I}_T}{T} \ \mathbf{R}^{\mathrm{T}} \mathbf{z}^{-1} \right ] + \ldots
\end{eqnarray}
where in the last line the identity $T \times T$ matrix has been systematically inserted between $\mathbf{R}$ and $\mathbf{R}^{\mathrm{T}}$ matrices, whenever they appear. The $\mathbf{z}^{-1}$ matrices in equation \eqref{Greenexp}, being multiples of the identity matrix, could be safely pulled out of the expectation map (which is to be meant with respect to the probability measure in \eqref{probmeas}). So, in order to compute a generic matrix element of the Green's function $\mathbf{g}_{\mathbf{c}}$, one would need to compute $n$-point correlation functions of the following kind: $\mathbb{E} \left [ R_{i_1 t_1} 
R_{i_2 t_2} \ldots R_{i_{2n} t_{2n}} \right ]$. Following \cite{Burda2004_1}, two different kinds of indices have been considered in this expression: indices of the $N$-type (ranging from $1$ to $N$) and indices of the $T$-type (ranging from $1$ to $T$). Moreover, an even number of matrix elements has been considered, since the Gaussian probability measure \eqref{probmeas} ensures the vanishing of all odd moments. Also, Wick's Theorem allows us to split any $n$-point correlation function into the sum of all possible products of two-point correlation functions (or propagators). For example, the four-point correlation function would read

\begin{equation} \label{4point}
\mathbb{E} \left [ R_{i_1 t_1} R_{i_2 t_2} R_{i_3 t_3} R_{i_4 t_4} \right ] = \mathbb{E} \left [ R_{i_1 t_1} R_{i_2 t_2} \right ] \mathbb{E} \left [ R_{i_3 t_3} R_{i_4 t_4} \right ] + \mathbb{E} \left [ R_{i_1 t_1} R_{i_3 t_3} \right ] \mathbb{E} \left [ R_{i_2 t_2} R_{i_4 t_4} \right ] + \mathbb{E} \left [ R_{i_1 t_1} R_{i_4 t_4} \right ] \mathbb{E} \left [ R_{i_2 t_2} R_{i_3 t_3} \right ],
\end{equation}
where the two-point correlation function is as in equation \eqref{notimecov}:

\begin{equation} \label{2point}
\mathbb{E} \left [ R_{it} R_{jt^{\prime}} \right ] = C_{ij} \delta_{t t^{\prime}}.
\end{equation}
Let us now represent the main ingredients in the Green's function expansion as follows (see Figure \ref{Symbols}). $N$-type indices will be represented as black circles, while $T$-type indices will be represented as grey circles. A $\mathbf{z}^{-1}$ matrix element will be represented as a straight solid line connecting two $N$-type indices, while elements of the $\mathbf{I}_T / T$ matrix will be represented as a dashed line connecting two $T$-type indices. The propagator in equation \eqref{2point}, in turn, will be depicted as a double arc connecting two pairs of indices, each made of an $N$-type and a $T$-type index.
With the previous positions, the first few terms in the expansion of the Green's function look like in Figure \ref{GreenDiag}. The Green's function is represented as a grey circle within $N$-type indices, while the other diagrams correspond to the different contributions in the expansion in \eqref{Greenexp}. As can be seen, such diagrams are divided into two categories: those with crossing lines (such as the one in the fourth line of Figure \ref{GreenDiag}) and those without crossing lines, also known as planar diagrams. It can be shown that the contribution of diagrams belonging to the former group vanishes in the infinite matrix limit \eqref{thermlim}. Intuitively speaking, this is because in planar diagrams closed loops (which give a contribution of order $N$ for black lines and a contribution of order $T$ for grey lines) and external horizontal lines (giving contributions of order $1/N$ for black lines and of order $1/T$ for gray lines) occur in equal numbers. So, in the thermodynamic limit the two contributions balance each other. On the other hand, non planar diagrams have extra $1/N$ factors which are not compensated by closed loops.
So, in the thermodynamic limit, the Green's function is only composed of planar diagrams, whose building blocks are horizontal lines and ``rainbow-like'' structures, as it is easily seen from Figure \ref{GreenDiag}. More formally, such rainbow diagrams are usually called one-line-irreducible (1LI) since they cannot be split into two parts by cutting one horizontal line (either solid or dashed). It is then convenient to introduce the self-energy $\mathbf{\Sigma}_{\mathbf{c}}$, \emph{i.e.} the generating function for such diagrams (see Figure \ref{SelfEnergy}). Then, the Green's function can be expanded in terms of the self-energy, as in Figure \ref{GreenS}.
Translating diagrams into equations, the meaning of Figure \ref{GreenS} is the following:

\begin{equation} \label{greensigmaexp}
\mathbf{g}_{\mathbf{c}}(z) = \mathbf{z}^{-1} + \mathbf{z}^{-1} \mathbf{\Sigma}_{\mathbf{c}} (z) \ \mathbf{z}^{-1} + 
\mathbf{z}^{-1} \mathbf{\Sigma}_{\mathbf{c}}(z) \ \mathbf{z}^{-1} \mathbf{\Sigma}_{\mathbf{c}}(z) \ \mathbf{z}^{-1} + \ldots =
\left ( \mathbf{z} - \mathbf{\Sigma}_{\mathbf{c}}(z) \right )^{-1}.
\end{equation}
Thus, the self-energy somehow represents an ``effective'' matrix that replaces the $\mathbf{c}$ matrix by removing the expectation map in the Green's function expansion of equation \eqref{Greenexp}. \\
Parallel to the previous definitions, it is convenient to introduce a second Green's function for the matrix $\mathbf{R}^{\mathrm{T}} \mathbf{R}$, which has exactly the same eigenvalues of the matrix $\mathbf{c} = \mathbf{R} \mathbf{R}^{\mathrm{T}}$ (plus some possible additional zero modes, depending on the relative size of $N$ and $T$). Let us define such Green's function as

\begin{equation} \label{Greenstar}
\widetilde{\mathbf{g}}_{\mathbf{c}} (z) = \mathbb{E} \left [ \left (T \ \mathbf{I}_T - z^{-1} \mathbf{R}^{\mathrm{T}} \mathbf{R} \right )^{-1} \right ],
\end{equation} 
in such a way that $\widetilde{\mathbf{g}}_{\mathbf{c}}$ would be represented by a diagrammatic expansion in which $N$-type and $T$-type indices (and consequently also dashed and solid lines) would switch roles with respect to the case of $\mathbf{g}_{\mathbf{c}}$, represented in Figure \ref{GreenDiag}. Also, by defining a new self-energy function $\widetilde{\mathbf{\Sigma}}_{\mathbf{c}}$ as the generating function for 1LI diagrams with switched indices and lines (with respect to the 1LI diagrams generated by $\mathbf{\Sigma}_{\mathbf{c}}$), it is possible to write the following relation

\begin{equation} \label{greensigmastar}
\widetilde{\mathbf{g}}_{\mathbf{c}}(z) = \left (T \ \mathbf{I}_T - \widetilde{\mathbf{\Sigma}}_{\mathbf{c}}(z) \right )^{-1},
\end{equation}
in complete analogy with equation \eqref{greensigmaexp}. \\
As already stated, in the thermodynamic limit Green's functions are only composed of planar diagrams and horizontal lines. On the other hand, all 1LI diagrams can be obtained by adding a propagator to some proper planar diagram (see Figures \ref{GreenDiag} and \ref{SelfEnergy}). This observation allows to establish two relations between the Green's functions and the self-energy functions (which contain all possible 1LI diagrams). Recalling the form of propagators (see equation \eqref{2point}), such relations can be written as

\begin{eqnarray} \label{DS2}
\mathbf{\mathbf{\Sigma}}(z) &=& \mathbf{C} \ \mathrm{Tr} \left [ \widetilde{\mathbf{g}}_{\mathbf{c}} (z) \right ]
\\ \nonumber
\widetilde{\mathbf{\Sigma}}(z) &=& \mathbf{I}_T \ \mathrm{Tr} \left [ \mathbf{C} \ \mathbf{g}_{\mathbf{c}}(z) \right ].
\end{eqnarray}
Equations \eqref{greensigmaexp}, \eqref{greensigmastar} and \eqref{DS2} form the so called set of Dyson-Schwinger equations, which can be solved for $\mathbf{g}_{\mathbf{c}}$ by consecutively eliminating $\mathbf{\Sigma}_{\mathbf{c}}$, $\widetilde{\mathbf{g}}_{\mathbf{c}}$ and $\widetilde{\mathbf{\Sigma}}_{\mathbf{c}}$. This yields

\begin{equation} \label{gG}
z \mathbf{g}_{\mathbf{c}}(z) = Z \mathbf{G}_{\mathbf{C}}(Z) \ , \ \ \ Z = \frac{z}{\mathrm{Tr} \left [ \mathbf{g}_{\mathbf{c}} (z) \right ]},
\end{equation}
where $\mathbf{G}_{\mathbf{C}}$ is as in equation \eqref{Green}. By carrying out all calculations explicitly one finds

\begin{equation} \label{zZ}
Z = \frac{z}{1 + q \left ( N^{-1} \ \mathrm{Tr} \left [ z \ \mathbf{g}_{\mathbf{c}}(z) \right ] - 1 \right )}.
\end{equation}
Recalling the definition of the moment generating functions (see equations \eqref{mgf} and \eqref{normtrace}), one can see that the relations in \eqref{gG} and \eqref{zZ} complete the derivation of equations \eqref{mMrel} and \eqref{confmap}, which was the goal of this Appendix. \\
Eventually, let us mention (as already pointed out in \cite{Burda2004_1}) that in the limit of very large samples, \emph{i.e.} when $T \rightarrow \infty$ with $N$ fixed, one has $q = 0$ and consequently equations \eqref{gG} and \eqref{zZ} yield $\mathbf{g}_{\mathbf{c}}(z) = \mathbf{G}_{\mathbf{C}}(z)$. This, of course, would cause the corresponding spectral densities to be identical, giving a rigorous meaning to the intuitive statement that in the limit of a large number of observations the eigenvalue spectrum of the $\mathbf{C}$ matrix is faithfully reproduced by its estimator $\mathbf{c}$.

\end{document}